\documentclass[11pt,preprint]{aastex}

\shorttitle{Turbulent Convection}
\shortauthors{Arnett, Meakin \& Young}

\newcommand{\lcz}{\ell_{CZ}}
\newcommand{\ld}{\ell_{D}}
\newcommand{\dV}{d{\cal V}}
\newcommand{\ubar}{(u')_{rms}}
\newcommand{\vrms}{v_{rms}}
\newcommand{\Tbar}{(T')_{rms}}
\newcommand{\urms}{\langle\overline{u'^{2}}\rangle^{1 \over 2} }
\newcommand{\Trms}{\langle\overline{T'^{2}}\rangle^{1 \over 2} }

\newcommand{\sound}{\cal C}

\newcommand{\etal} { et~al.\ }  

\def
\nuc#1#2{\relax\ifmmode{}^{#1}{\protect\text{#2}}\else${}^{#1}$#2\fi}

\def \avg#1{\relax \overline{\langle #1 \rangle}}
\def \havg#1{\relax \langle #1 \rangle}

\begin{document}

\title{Turbulent Convection in Stellar Interiors. II. The Velocity
Field}

\author{David Arnett\altaffilmark{1},
 Casey Meakin\altaffilmark{1,2,4}, and
 Patrick A. Young\altaffilmark{1,3}
\altaffiltext{1}{Steward Observatory, University of Arizona, 
933 N. Cherry Avenue, Tucson AZ 85721}
\altaffiltext{2}{FLASH Center, University of Chicago,
Chicago, IL}
\altaffiltext{3}{School of Earth and Space Exploration, Arizona State 
University, Tempe, AZ}
\altaffiltext{4}{Joint Institute for Nuclear Astrophysics, University 
of Chicago, Chicago, IL}
\email{ darnett@as.arizona.edu,casey.meakin@gmail.com,
patrick.young.1@asu.edu}
}

\begin{abstract}
We analyze stellar convection with the aid of 3D hydrodynamic 
simulations, introducing the turbulent cascade into our theoretical
analysis. We devise closures of the Reynolds-decomposed mean field
equations by simple physical modeling of the simulations (we
relate temperature and density fluctuations via coefficients); the
procedure (CABS, Convection Algorithm Based on Simulations) is 
terrestrially testable and is amenable to systematic improvement.
We develop a turbulent kinetic energy equation which contains both
nonlocal and time dependent terms, and is appropriate if the convective 
transit time is shorter than the evolutionary time scale.
The interpretation of mixing-length theory (MLT) as generally
used in astrophysics is incorrect; MLT forces the mixing length to
be an imposed constant. Direct tests show that the
damping associated with the flow is that suggested
by Kolmogorov ($\varepsilon_K \approx \rho \ubar^{3}/\ld$, 
where $\ld$ is the size of the largest eddy and $\ubar$ is 
the local rms turbulent velocity). This eddy size is approximately the depth of 
the convection zone $\lcz$ in our simulations, and corresponds in some respects 
to the mixing length of MLT. New terms involving the local heating 
due to turbulent dissipation should appear in the stellar evolutionary
equations, and are not guaranteed to be negligible.
The enthalpy flux (stellar ``convective luminosity'') is directly connected to
the buoyant acceleration, and hence to the scale of convective velocity.
MLT tends to systematically underestimate the velocity scale,
which affects estimates of chromospheric and coronal heating, 
mass loss, and wave generation.
Quantitative comparison with a variety of 3D simulations reveals 
a previously unrecognized consistency.
Extension of this approach to deal with rotational shear and MHD is 
indicated. 
Examples of application to stellar evolution will be presented
in subsequent papers in this series.
\end{abstract}

\keywords{stars: evolution - hydrodynamics - convection -
turbulence}

\section{Introduction}

More than fifty years ago,  the version of the
``mixing length theory'' of convection (MLT)
which became the preferred basis for subsequent study of stellar
evolution
was introduced \citep{bier51,vitense53,bv58}. 
Despite much effort
(still ongoing), MLT is still the standard choice for the field. 

In this paper we develop a new procedure, "Convective Algorithms
Based on Simulations" (CABS), in which we close the 
Reynolds-decomposed, angular and time averaged equations by simple 
physical models based upon analysis of fully three-dimensional, time 
dependent turbulent stellar convection \citep{ma07b}. These
simulations include convective boundaries within the computational
volume (as far from the edges of the grid as feasible), and allow
interface physics to be examined.
The resulting theoretical formalism allows us to incorporate content
from other simulations (especially 
\cite{cs89,cs96,kim95,kim96,pw00,pwj00,rob04}), and
from research in other fields, such as terrestrial fluids
\citep{turner73}, oceanography \citep{gill},
and meteorology \citep{dutton}, which have a firmer 
empirical basis. We develop a simple description of the convective
velocity field as seen in our simulations. This effort 
brings some startling suggestions for revision of our interpretation 
of MLT, and suggests how our approach may be generalized to include
rotation and magnetic fields \citep{balbus,pessah}.
This is timely, considering recent success in simulating turbulent 
plasma with magnetic fields \citep{brown08,schu08}.

Since the formulation of MLT, there has been a considerable
development
in understanding the nature of chaotic behavior in nonlinear systems;
see \cite{cvit} for a review and reprints of original papers,
and \cite{frisch,gleick,thomp}. 
\cite{lorenz} presented a solution to the Rayleigh problem
of thermal convection \citep{chandra} which captured the seed of
chaos in the Lorenz attractor, and contains a representation of the
fluctuating aspect of turbulence not present in MLT.
\cite{kolmg} and \cite{obukhov} developed the modern version of the
turbulent cascade. Although already formulated, the original theory 
\citep{kolmg41} was not used in MLT.

We derive our approximate theory from a consideration of the full
equations of 3D compressible hydrodynamics for a multiple component
fluid. These are close to the corresponding equations for a high beta
(matter dominated) plasma.
This approach will allow us to incorporate a variety of 
phenomena in a coherent way, and to evaluate their relative 
importance, rather than to patch together various bits of physics
piecemeal. Here we focus on the dominant features of
non-rotating, non-magnetic, turbulent, compressible fluid flow. The
results are applicable to almost all stages of stellar
evolution (any stage having convection or shear)\footnote{The shear 
from convection is similar to the shear from differential 
rotation; fluid experiments may use either to investigate the physics 
of shear \citep{turner73}. Although different in some details, 
there are deep connections between 
convective mixing as described here and the rotational mixing 
investigated by \cite{mm00}.}.

In Section~2 we examine the physical aspects of stellar convection, 
and show  that the velocity scale is set by the balance between 
buoyant driving, and damping in the Kolmogorov cascade. 
Appropriate averaging allows us to deal with mild time dependence
and reveals a robust underlying behavior. 
We develop a kinetic energy equation describing the average
properties of turbulent convection, which shows how turbulent 
motion is created, transported, and destroyed.
In Section~3 we incorporate this theoretical development into
the equations of stellar evolution
with turbulent convection, including new terms.
Section~4 uses the theoretical development to compare our
simulations to others, and finds previously unrecognized similarities.
Section~5 indicates some important implications of this work,
including how
the effects of burning, rotation and magnetic fields may be included.
Section~6 summarizes our results and conclusions. 
Quantitative treatment of the dynamics
of fluctuations and a comprehensive algorithm for stellar 
evolutionary calculations will be developed in subsequent papers 
in this series.

\section{Physical Aspects of Stellar Convection}

Stellar convection has high Reynolds numbers because of the large
linear scales, and is therefore highly turbulent. Our simulations have
adequate resolution to show this type of behavior. Turbulent 
convection has several key features that need to be modeled:
(1) it is nonlocal, (2) it has strong fluctuations in both space and
time, (3) it is damped by a cascade of unstable vortices down to
scales small
enough for microscopic processes to dissipate effectively, (4) mixing
of passive scalar properties is efficient, (5) turbulent behavior 
spreads to fill the volume allowed, and (6) buoyant acceleration is
closely related to convective (enthalpy) flux, so that 
convective kinetic energy
is closely tied to convective luminosity. MLT incorporates (4) and
imperfectly deals with (6); we will address all six issues.

\cite{catt91} found that 3D simulations of turbulent convection had
two aspects:
(1) vigorous, large scale downflows\footnote{\cite{pw00} suggest that
large scale flows do dominate the energy flux.}, 
and (2) disorganized weaker motions. These two aspects have been 
confirmed by many other 
simulations,  including our own. MLT attempts to describe the average
properties of the disordered 
aspect. We will construct a theory which includes both; buoyant 
acceleration is characteristic of the largest scales, and turbulent 
dissipation of the smallest.

\subsection{The Kinetic Energy Equation}

We start with the equations we used in our 3D simulations
\citep{ma07b}.
We use Reynolds decomposition of relevant flow and 
thermodynamic variables, which separates the fluctuating from the
slowly varying components \citep{tennekes}. 
Numerical simulations
dissipate features with wavelengths at or below the grid scale, 
and we will identify this with dissipation in the turbulent cascade
\citep{kolmg41,kolmg}. 

In the process of approximation of partial differential equations by
finite
methods, there is inevitably a loss of information at scales smaller
than the grid size. A single volume element in space is approximated
as a homogeneous entity; this is equivalent to complete mixing at this
scale, at each time step, of mass, momentum, and energy. The loss
of information that occurs with this mixing corresponds to an 
increase in entropy \citep{shannon}; the mixing of  momentum is 
equivalent to the action of viscosity \citep{ll59}. In 3D flow,
turbulent 
energy will cascade from large
scales to small, at a rate set by the largest scales. 
We use an implicit sub-grid dissipation in our large eddy 
simulation (ILES), which is the most computationally
efficient way to deal with turbulent systems with a large range of
scales
\citep{boris07,wood07}; the largest scales, which set the rate of
cascade and contain most of the energy, are resolved on our grid and
explicitly calculated, while the 
sub-grid scales are dissipated in a way consistent with the Kolmogorov
cascade.

\cite{sytine} demonstrated
that PPM, the piece-wise parabolic method based on the Euler equation
(which has no explicit
viscosity), converges to the same limit as methods based on 
compressible viscous equations  (which do have explicit viscosity),
as the grid is refined to
smaller zones and smaller effective viscosity (the relevant limit
for astrophysics). 
\cite{porter99} show the compatibility of
mildly compressible flow with the \cite{kolmg41} spectrum;
\cite{knpw07} have pushed this to highly compressible flows as well.
To represent the sub-grid dissipation, which
is inherent in our simulations, we explicitly introduce a volumetric
dissipation rate for kinetic energy, $\varepsilon_{K}$, in our 
theoretical analysis. However, the turbulent cascade is a property
of the whole convective flow, so connection to turbulence theory must 
be made through {\em integrals} of  $\varepsilon_{K}$ over the 
convection zone. Note that this is different from  defining
 $\varepsilon_{K}$ as a function of {\em local} variables (e.g.,  
 \cite{smag63,cs89,hansen,sa00}); see below.

The kinetic energy (KE) equation for convective motion
was given in \cite{ma07b}. Taking the scalar
product of the velocity with the equation of motion, we decompose the
convective velocity $\bf u$, the density $\rho$, and the pressure $p$ 
into mean and fluctuating components (e.g., $p=p_{0}+p'$, so
the time averages are
$\overline{p} = p_{0}$ and $ \overline{p'} =0$.
This choice of just $\bf u$, 
$p$, and $\rho$ for this Reynolds decomposition
into average and fluctuating parts gives the simplest equation for 
kinetic energy; the velocity $\bf u $ 
is derived from buoyancy and pressure forces ($\rho$ and $p$ fluctuations). 

Using the hydrostatic
equilibrium condition, and performing averages, gives 
(see eq.~A.12 in \cite{ma07b}),
\begin{eqnarray}
\partial_t \langle\overline{\rho E_K}\rangle 
+ \nabla \cdot \langle\overline{\rho E_K {\bf u_0}}\rangle
= \nonumber\\
- \nabla \cdot \langle\overline{{\bf F_p + F_K }}\rangle
+ \langle\overline{ p' \nabla \cdot u' }\rangle
\nonumber\\
+ \langle\overline{ \rho' {\bf g \cdot u'} }\rangle
- \varepsilon_K .
\label{a12}
\end{eqnarray}
We use $\langle p \rangle$ 
to denote an average over angles at constant radius.
The time average is taken over durations greater than a transit time
$t_{transit} =  \lcz /\ubar$, where $\lcz$ is the depth of 
the convective region, and $\ubar$ is the rms velocity across this
region.
This smooths the fluctuations, gives a nonlocal character to the
analysis, and
implies a separation of time scales into short ($t \ll t_{transit}$)
and long ($t \gg t_{transit}$). 
We consider the case in which we may integrate over these short time
scales and explicitly calculate the evolution on the long time scales.

Here $E_{K}$ is the kinetic energy per unit mass, $\rho$ the mass 
density, $\bf u_{0}$ is the nonfluctuating part of the fluid velocity
vector and $\bf u'$ its fluctuating part, so
\begin{equation}
{\bf F}_p =  p' {\bf u'}  
\end{equation}
is the energy flux due to pressure perturbations carried by 
fluctuations (this pressure-velocity correlation flux reduces to the 
acoustic flux when considering sound waves, \cite{ll59},
p.~251, and contains the energy flux due to internal gravity waves),
\begin{equation}
{\bf F}_K =  \rho E_K {\bf u'}  
\end{equation}
is the flux of kinetic energy carried by convective turbulent motion,
$\bf g$ is the gravitational acceleration vector, and
$\varepsilon_{K}$
is the volumetric rate of dissipation implied by the turbulent cascade
down to small scales.

\begin{figure}
\figurenum{1}
\plotone{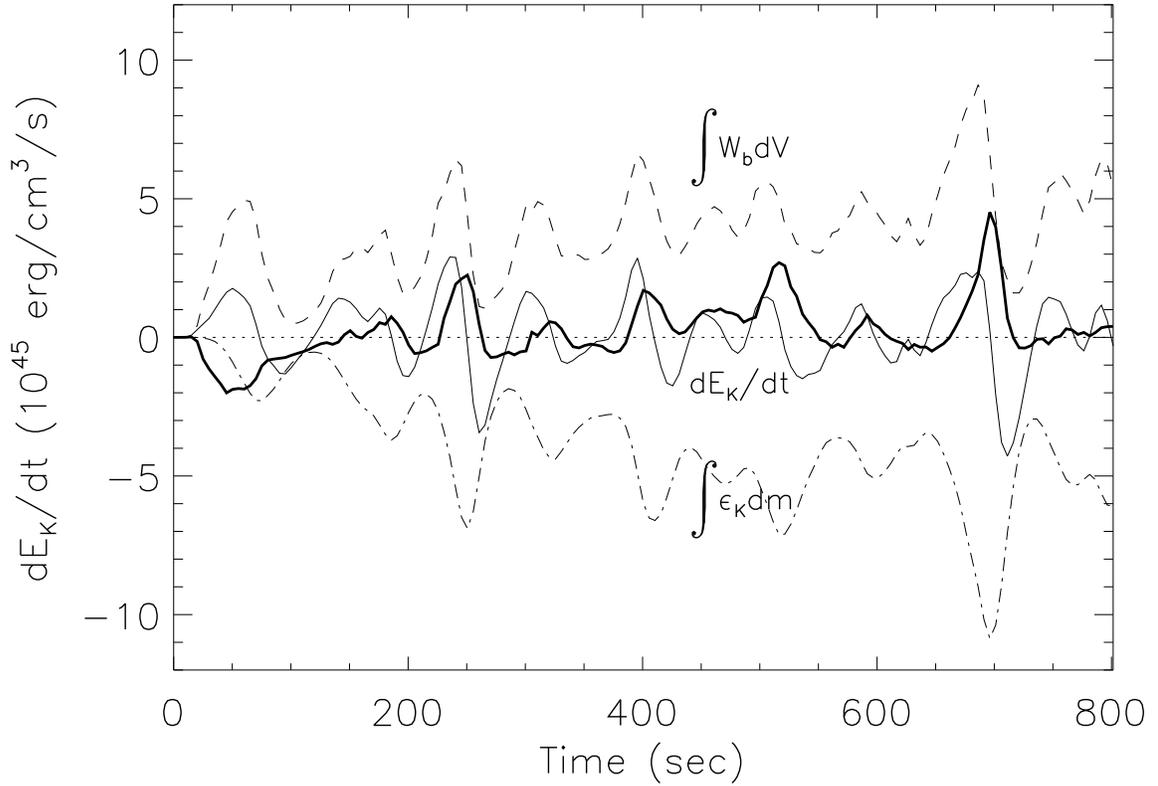}
\caption{Behavior of global quantities in the convection zone, which
affect the turbulent kinetic energy, entrainment, and gravity wave
generation at the convective boundaries. The thick line is the sum
of all terms except entrainment and boundary wave luminosity, and 
indicates when entrainment events are vigorous (\cite{ma07b}, Fig.~4).
The buoyancy driving $B= \int W_{b}\dV$ is compensated for by the 
turbulent damping $D= \int \varepsilon_{K} \dV$. 
The time derivative of the kinetic energy in the convection zone is 
denoted $dE_{K}/dt = { d \over dt} \int \rho \ubar ^{2} \dV$.
There is a slow increase in amplitude due to lack of balance between
nuclear heating and neutrino cooling (see Fig.~\ref{fig5} below and
\cite{arnett}).
}
\label{fig3a}
\end{figure}

\subsection{Boundaries}

The location of convective boundaries is a long-standing 
problem in stellar astrophysics. It has long
been known that some sort of "convective overshooting" is
necessary for models to match observations. This has been conceived
as mixing of material beyond the convective boundary, often by
"penetrative convection". This problem exists in part because the
convective boundary has been inappropriately defined (\cite{ma07b},
especially \S4.1 and \S7). The standard
definition used in mixing length theory defines the convective
boundaries using the thermodynamic nablas, which essentially mark the
onset and cessation of buoyant acceleration. 
One of the essential properties of convection is that turbulence
fills the space available. A more appropriate criterion is one in 
which the stellar background is stiff enough to contain the 
turbulent convection, and therefore must account for the relative 
strength of the turbulent flow and the elastic stable layers.

We will define the
convective zone as that region in which the stratification
of the medium is unstable to turbulent mixing. This is evaluated with
the "bulk Richardson number" 
\begin{equation}
Ri_B = \Delta b \ l/u^2,
\end{equation}
where $ \Delta b = \int_{\Delta r}¥ N^{2} dr$ is the buoyancy jump across a layer
of thickness $\Delta r$ in the radial direction,
$N^2$ is the Br\"{u}nt-V\"{a}is\"{a}l\"{a} frequency, $u$ is
the rms velocity providing shear, and $l$ is the scale length of the turbulence
(essentially the size of the largest eddies). 
The precise definition of the thickness $\Delta r$ is a topic 
deserving further study; in several cases we have noticed that the 
rapid change in $N^2$ near a boundary tends to make $\Delta b$ 
insensitive to the exact value of $\Delta r$ chosen for integration to
obtain $\Delta b$.

In the linear limit of plane parallel flow, a region
with $Ri_{g} \la 0.25$ has enough kinetic energy to overcome the
stable stratification \citep{dutton}. Here $Ri_{g}$ is the 
``gradient'' Richardson number, and is a locally defined quantity, 
used in a linear stability analysis. The ``bulk'' Richardson number 
$Ri_{B}$ is an inherently nonlinear quantity based on integration 
over an extended region. 
Since $Ri_{g,crit} > 0$, this allows for
$N^2 > 0$ as well. This formulation recognizes that layers that are
thermodynamically stable (real roots to $N^2$) can be
hydrodynamically unstable when kinetic energy is input to the zone.
The bulk Richardson criterion allows more mixing than predicted by
the Ledoux criterion; the Schwarzschild criterion ignores compositional
effects, and also predicts more mixing than Ledoux (see 
\cite{kippen,hansen}.

Stable regions adjacent to the convective zone will become Richardson
unstable periodically as shear builds up from adjacent turbulent 
motions and waves generated by convection, leading to entrainment of
material at convective boundaries \citep{fernando}. The bulk Richardson 
number is not only 
an indicator of a convective boundary; it also determines the 
``rate'' at which the boundary migrates through the lagrangian mesh
by mixing processes \citep{ma07b}, and thus provides a ``dynamic'' 
boundary condition for the flow.

\subsection{Averages}

Our analysis is based primarily upon the numerical simulations of 
oxygen shell burning in a $23 \rm M_{\odot}$ star, but is found to 
be broadly applicable to other examples of stellar convection 
(such as convective envelopes, which are driven by the superadiabatic
layer at the top rather than the nuclear burning shell at the bottom). 
Our convective zone has a depth of two pressure scale heights 
(see \cite{ma07b} for detail).

Why do we need to do averaging? Fig.~\ref{fig3a} shows the behavior of
the largest terms in the kinetic energy equation (averaged over the 
convection zone), throughout the duration of the simulation. The 
dominant terms are the integrals of the buoyancy driving 
$\int W_b \dV = \int g \rho' u' \dV $, 
which is positive, and of
damping, $\int \varepsilon_K \dV$, which is negative. Both show
recurrent bursts (ten in 800 seconds). The time derivative
of kinetic energy in the convection zone (labeled $dE_K/dt$)
is constant on average. The remaining bumps are due to missing terms 
(divergence of $F_P$ and $F_K$).

We saved complete data every 0.5 seconds,
which was adequate for making movies, and constructing averages of
motion. A key issue is distinguishing between turbulent velocities
and wave velocities with such a coarse stride in time. 
We have found a way to split the velocity averages 
into turbulent and wave components for better accuracy in the analysis
(see Appendix~A for a detailed discussion of how this is done).

\begin{deluxetable}{rrrr}
\tablecaption{Kinetic Energy Equation Terms Averaged
over Convection Zone  and over four
transit times ($\rm erg/s$) \label{tablex}}
\tabletypesize{\small}
\tablewidth{300pt}
\tablehead{\colhead{Term} & \colhead{Value} &\colhead{Term} & 
\colhead{Value} }
\startdata
$\int \langle \overline{ W_{B} } \rangle d{\cal V}$ & 4.576(45) &
$-\int \langle \overline{\varepsilon_{K}} \rangle d{\cal V}$ &
-4.677(45) \\
$-\int \langle \overline{ \nabla \cdot F_{K} } \rangle d{\cal V}$ &
2.584(44) &
$-\int \langle \overline{ \nabla \cdot F_{P} } \rangle d{\cal V} $ &
-9.922(43) \\
$-\int \langle \overline{ dE_{K}/dt } \rangle d{\cal V} $ &
-5.790(43) &
$\int \langle \overline{p' \nabla u'} \rangle d{\cal V} $ &
-1.516(41) \\
\enddata
\end{deluxetable}

The level of complexity shown in Fig.~\ref{fig3a} is daunting, but 
the near balance of buoyant driving and turbulent damping provides 
a clue.

The system is simplified if we integrate over time; the
resulting values over four transit times are shown in 
Table~\ref{tablex}. We do not calculate
the damping directly, but deduce it as the remainder left from the 
other five terms, so that the entries sum to zero. We are comfortable
with this procedure because of the good conservation properties of the
numerical simulations.

The dominance of
buoyant driving and damping is now clear; the next largest term is 
the divergence of the kinetic energy flux, at less than 6 percent
of the buoyancy term.

Suppose the globally-averaged damping term has the Kolmogorov form, 
so that
\begin{equation}
 \int_{CZ} \varepsilon_K  d {\cal V} 
  = M_{CZ} {\vrms}^{3}/ \ld ,
\label{kolgeps}
\end{equation}
where $\ld$ is a constant "damping length," and $\vrms$ is the
rms velocity determined over the entire convection zone. 
For a turbulent spectrum, $\ld$ is 
taken to be the largest length scale \citep{ll59}. 
In what follows we will use the term Kolmogorov damping to mean
damping due to a turbulent cascade to small scales, having this cubic
dependence on rms velocity.

Unlike MLT, we relate the damping length to the {\em largest} scale
in the flow; this may be thought of as a ``coherent structure''
or a ``plume which traverses the depth of the convection zone.''
It is not a free parameter. We introduce the notation 
\begin{equation}
    \alpha_{D}= \ell_{D}/\ell_{CZ},
    \label{eqalphad}
\end{equation}
so that $\alpha_{D}$ would be constant if the size of the largest eddy
scales simply with the depth of the convection zone.
\cite{knpw07} found the Kolmogorov damping to be very close to
constant during a statistical steady state in their 3D ($2048^{3}$)
simulations.

\begin{deluxetable}{rrr}
\tablecaption{Time Scales \label{tablet}}
\tabletypesize{\small}
\tablewidth{300pt}
\tablehead{\colhead{Time scale} & \colhead{Definition} &\colhead{Value(seconds)} 
}
\startdata
Buoyant rise & $\int W_b \dV/KE $ & 14.1 \\
Velocity damping & $\ld / \vrms $ & 27.6 \\
KE damping & $\ld / (2 \vrms)$ & 13.8 \\
Transit time & $\lcz / \vrms $ & 51.4 \\
Turnover time & $2\lcz /  \vrms$ & 102.8 \\
\enddata

\end{deluxetable}

The mass contained in the convection zone is $M_{CZ} = 1.84 \times 
10^{33} \rm\ g$ and the total kinetic energy is $ \int E_{K} d{\cal V} = 
8.61 \times 10^{46} \rm\ ergs$ so that the rms velocity is $\vrms = 
9.66 \times 10^{6} \rm\ cm/s$. This 
gives $ \ld = 3.56 \times 10^{8}{\rm\ cm} \approx 0.85\ 
\lcz$. It is natural to compare the damping length to the depth of the
convection zone. Not only is $\lcz$ the largest length available 
for an eddy, but if measured in pressure scale height units
($ \ell_{CZ}/H_{P}$), it indicates
the degree of thermodynamic anisotropy across the convective region. 

Table~\ref{tablet} gives several relevant time scales in seconds. Although
these times are short in human terms, they are much closer to thermal
relaxation times than are the corresponding numbers for deep simulations 
of the solar convection zone (our simulations are much more relaxed in this
sense).

We may write the global damping as
$   M_{cz} v_{rms}^{3}/ \ld =   {1 \over 2}M_{cz} v_{rms}^{2}/ (\ld 
/2 \vrms )$, and see that the damping time for kinetic energy in the
convective zone is half the time to transit a damping length (which
is approximately the depth of the convection zone).  
The {\em turnover} time for the convection zone 
is $2\tau$, where $\tau = \lcz/\vrms$ 
is the transit time, to be precise. The rise time for kinetic energy
and the corresponding damping time are similar (14 seconds), and
much shorter than a turnover time.

The term ``convective efficiency'' in the stellar context is usually 
taken to mean that the time scale for convective energy transport is
much less than the time scale for radiative energy transport 
(\cite{hansen}, p.~187);
this insures that the actual temperature gradient is only slightly
in excess of the adiabatic one. Convection can also be thought of
as a thermodynamic cycle, taking a time $2 \lcz/ \vrms $.
As we saw above, the dissipation time scale for kinetic energy is 
about one-seventh of this (0.134), so that in this sense, convection 
is not thermodynamically efficient at all, but requires continual work to
keep it running.
The two uses of ``efficiency'' causes confusion, and downplays the
fact that stellar convection is {\em highly} dissipative, even for slightly 
superadiabatic temperature gradients.

\begin{figure}
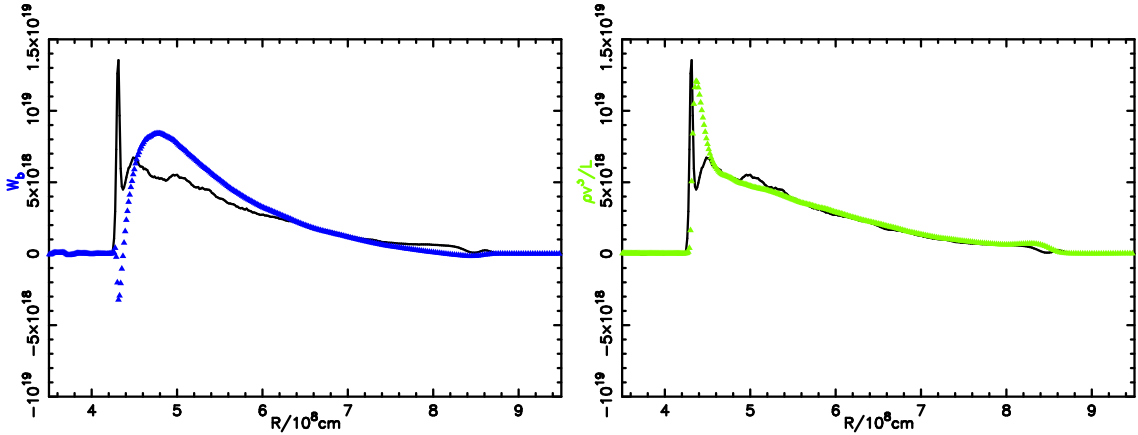

\figurenum{2}
\includegraphics[angle=-90,scale=0.3]{f2a.eps}
\includegraphics[angle=-90,scale=0.3]{f2b.eps}
\caption{Comparison of terms in the kinetic energy equation, averaged
over four transit times and over angle, as a function of radius. The solid 
line is the inferred local value of the volumetric dissipation due to
the turbulent cascade. The thick line of triangles represents a term
in the equation for kinetic energy density (Eq.~\ref{a12}).
Buoyancy driving (left) and $\varepsilon_K$ approximated as 
$\rho u_{rms}^{3} / \ell_{D} = 1.54 \rho u_{t}^3/\ld$
(right) are shown. 
Here we distinguish between the global rms convective velocity 
$\vrms$ and the rms convective velocity at a given radius, 
$\urms = \ubar$. We find $\varepsilon_{K}= (\vrms )^{3}/\ld$ for 
$\ld=3.5 \times 10^{8} \rm cm \approx  \ell_{CZ}$, {\em globally}, 
and with this choice of damping length, the {\em local} damping 
per unit mass is proportional to $u_{rms}^3/ \ld$. Much of the variation
shown in the bottom panel is simply due to the density gradient (see
the angular velocities in \cite{ma07b}, Fig.~6, left panel, and the
background density structure in their Fig.~2, top left panel).
Note that, as shown,
the signs of the terms are consistent with the kinetic energy
equation, so that up means increasing, down decreasing kinetic 
energy for the terms indicated by triangles.
}
\label{fig99a}
\end{figure}

\subsection{Anisotropy and Kinetic Energy\label{catt}}

Here we present an initial, qualitative discussion of the structure 
of the flow, which 
we are investigating quantitatively for subsequent publication.
We extend the idea of \cite{catt91} that the convective velocity
field has two components, a more ordered global flow and a chaotic
turbulent flow. We focus on the source and sink for convective
kinetic energy. Gravitational acceleration breaks the symmetry
of space; we choose our $z$-axis parallel to this acceleration. 
Buoyant acceleration starts an anisotropic flow in the $z$ 
direction.  The flow is unstable, and begins to break up into smaller
scales, and becoming more isotropic.

Suppose that the flow occurs in narrow, vigorous down plumes and 
slow wide upflows
for convective zones with significant anisotropy. The kinetic energy
in such a downflow would dominate that in the upflow; the kinetic energy
flux has two more powers of velocity than the mass flux, so that the 
fastest flows dominate the kinetic energy budget \citep{ma07b}. 
We separate the kinetic energy into two
components, $u_{a}^{2}/2$, which corresponds to the largest eddies and
$u_{b}^{2}/2$, which represents all of the turbulent cascade.
The turbulent instability begins to make the flow more isotropic, so that
\begin{equation}
    u_{z}^{2} \approx u_{a}^{2} + u_{b}^{2},
\end{equation}
in the $z$-direction (see Ch.~VI in \cite{batch}). If this 
kinetic energy is shared with the two perpendicular directions, then 
\begin{equation}
 u_{x}^{2} + u_{y}^{2} \approx u_{z}^{2}. 
\end{equation}
Further, it seems that the cascade component in the $z$-direction
is similar to the components in the $x$- or the $y$-directions, 
$ u_{x}^{2} \approx u_{y}^{2} \approx u_{b}^{2} $. While the
downflow matter is balanced by a return flow in $x$ and $y$,
the return flow is slower and broader (by our assumptions), with 
lower kinetic energy (which we neglect here). Thus
\begin{equation}
    u_{b}^{2} \approx u_{a}^{2}. 
\end{equation}
One quarter of the kinetic energy is in the largest scale component 
$u_{a}^{2}/2$. The ratio of vertical to horizontal
rms velocities is therefore $ u_{x}/u_{z} \approx 1/\sqrt{2}$.

How does this simple model compare with simulations?
In Fig.~6, \cite{ma07b} give
the rms values of velocity in the vertical ($v_{r}$) and the horizontal
($v_{\theta}$ and $v_{\phi}$) directions. We identify locally the
$r$ direction with the $z$-axis, and theta and phi with the $x$ and
the $y$ axes. Then from their Figure~6, we find 
\begin{equation}
    0.5 \leq u_{x}/u_{z} \approx u_{y}/u_{z} \leq 0.8,
\end{equation}
over the convection zone, away from the boundaries.
The measured value of the vertical velocity contains both large scale 
and cascade components. This is consistent with our estimate of 
$u_{x}/u_{z} \approx u_{y}/u_{z} \approx 1/\sqrt{2} \approx 0.707$. 
The large scale component of kinetic energy is comparable to the 
cascade one, consistent with \cite{catt91}, and our simulations. 

The simulations have led us to a simple, two-component model for the origin and
destruction of convective kinetic energy. Anisotropy is due to buoyant
acceleration in the largest eddy, which represents one component. 
The second is isotropic turbulence, which gives dissipation of kinetic 
energy after the cascade to small scales. It will be interesting to 
test and refine this model against a wide variety of simulations.
The power spectra of the flow will be a useful tool to explore this 
ansatz.

In what follows we will define a
turbulent rms velocity  $u_{t}$ from the measured horizontal 
velocity components,
$ u_{t}^{2} = {3 \over 2} ( u_{x}^{2} +u_{y}^{2} ) = {3 \over 4} 
u_{rms}^{2}$. We note for future reference that 
$ u_{rms}^{3}/u_{t}^{3} = ({4 \over 3})^{{3 \over 2}} = 1.540$.
Roughly speaking, we identify $u_{t}$ with the turbulent cascade.
The anisotropic component $u_{a}$ is identified predominately 
with the largest turbulent scale, and may be
sensitive to the structure of the convection zone and the location
and strength of the buoyant acceleration.

\subsection{Radial variation of Averages}

\begin{figure}
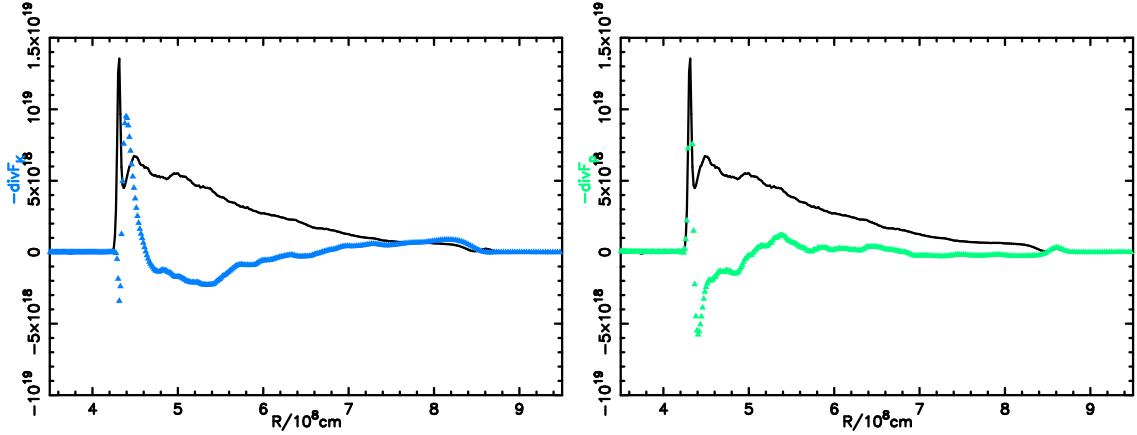

\figurenum{3}
\includegraphics[angle=-90,scale=0.3]{f3a.eps}
\includegraphics[angle=-90,scale=0.3]{f3b.eps}

\caption{Comparison of terms in the kinetic energy equation, averaged
over
four transit times and over angle, as a function of radius. The solid 
line is the inferred local value of the volumetric dissipation due to
the turbulent cascade. The thick line of triangles represents a term
in the equation for kinetic energy density (Eq.~\ref{a12}). Divergence
of $F_K$ (left) and divergence of $F_P$ (right) are shown. 
The terms for divergence of flux are important at the boundaries; 
they smooth the distribution of kinetic energy, causing the turbulent 
velocity to approach the form required for Kolmogorov damping.
}
\label{fig99b}
\end{figure}

\begin{figure}
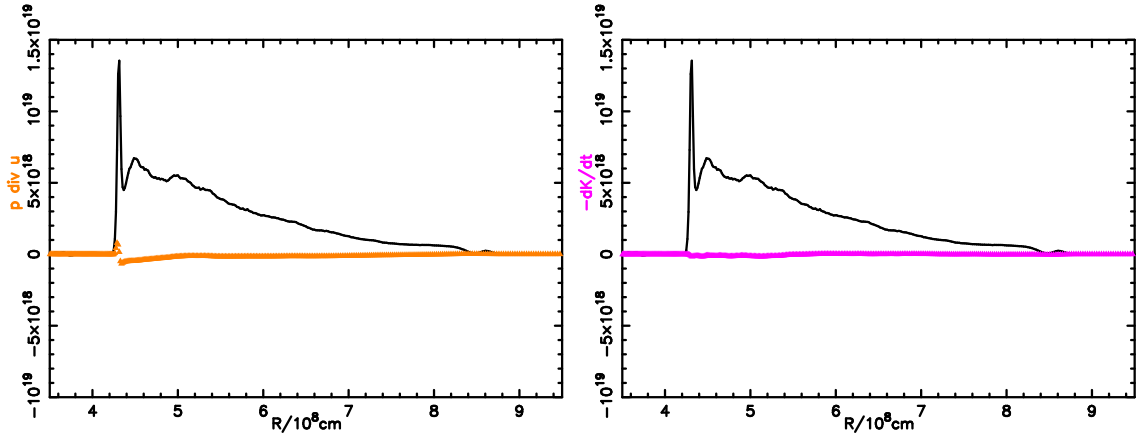

\figurenum{4}
\includegraphics[angle=-90,scale=0.3]{f4a.eps}
\includegraphics[angle=-90,scale=0.3]{f4b.eps}

\caption{Comparison of terms in the kinetic energy equation, averaged
over
four transit times and over angle, as a function of radius. The solid 
line is the inferred local value of the volumetric dissipation due to
the turbulent cascade. The thick line of triangles represents a term
in the equation for kinetic energy density (Eq.~\ref{a12}).
The terms $p' \nabla \cdot u'$ (left) and $dE_K/dt$ (right) are
shown. 
The effects of  $p' \nabla \cdot u'$ are small and most noticeable
at the lower boundary.
The change in the average kinetic energy is smaller still.}
\label{fig99c}
\end{figure}

We now relax one level of averaging, and consider the radial variation
of all the terms in the kinetic energy equation.
Figures~\ref{fig99a}, \ref{fig99b} and \ref{fig99c} show six terms
in six panels, each compared to the 
inferred damping (the solid line present in each panel). This allows us
to determine what each of five terms contribute to the inferred 
damping, and also shows an isotropic Kolmogorov estimate as a
simple approximation (Fig.~\ref{fig99a}, right panel). 
The solid line has a sharp spike at the bottom
of the convection zone. This feature is due to the dominance of g-mode
character in the motion near the  boundary region. The Kolmogorov 
cascade is appropriate away from the boundaries.

In Figure~\ref{fig99a}, the left panel shows the averaged buoyancy
$\int \overline{\langle W_B \dV \rangle }$, and the 
right panel shows the isotropic Kolmogorov approximation, both as
dotted curves. In planar geometry (a fair approximation), we can
relate this quantity to the "buoyancy flux" which is
widely used in experimental fluid mechanics \citep{turner73}.
The buoyancy flux is 
\begin{equation}
    {\bf q} = \avg{ {\bf g \cdot u' }\rho' } / \rho_{0}, 
\end{equation}
so that 
$ \int \overline{\langle W_B \dV \rangle }\rightarrow 
\int \overline{\langle q \rangle }4 \pi r^{2}dr $.
The enthalpy (convective heat) flux is
\begin{equation}
{\bf F_e} = \avg{ \rho_0 {\bf u'} C_p T' }, \label{fenthalpy}
\end{equation}
where $ C_p $ is the specific heat at constant pressure.
For low Mach number flows like ours\footnote{As we discuss below, 
simulations with stiffer equations of state will have larger pressure
fluctuations. We expect the neglected terms to be important for
wave generation, but probably in a restricted volume of the convection
zone.}
the pressure fluctuations are small.
The temperature and density perturbations at constant pressure
are proportional, 
\begin{equation}
    \rho'/\rho_0 = \beta_{T}\ T'/T_0, 
    \label{deltaeq}
\end{equation}
where $\beta_{T} = -\partial \ln \rho / \partial \ln T\Big )_{P}$, 
taken at constant pressure (and composition)\footnote{In \cite{ma07b} 
we denoted this quantity by $\beta_{T}$; the subscript $T$ is to 
avoid confusion both with Eddington's use of 
$\beta$ for the ratio of gas to total pressure, and the fluid dynamics
community use of $\beta$ as the adverse temperature gradient. }.  
We find, using
hydrostatic equilibrium of the unperturbed star $H_p g = P_0/\rho_0$,
\begin{equation}
{\bf F_e} =  \rho_0  H_p {\bf q} /\nabla_a. \label{fetoby}
\end{equation}
where $\nabla_a = \beta_{T}P/\rho C_{P}T $, a factor that we will see 
again. Except possibly for extreme cases, 
the enthalpy flux $F_{e}$, which is intimately related to the stellar
luminosity, is itself closely related to the buoyancy flux $q$ and
{\it hence to the convective velocity} (see also \cite{ma07b}). 
MLT ignores this connection; we shall exploit it.
The source of turbulent kinetic energy is directly proportional to
the convective luminosity and thus to the radial entropy gradient 
(``superadiabatic gradient'') of MLT.

Buoyancy is one of the  two dominant terms, 
and with a sign opposite to that of the damping; it differs in magnitude
from the inferred damping at the convective boundaries (lower) and
the broad peak above the bottom of the convection zone (higher).

The right panel shows the isotropic Kolmogorov approximation to the 
{\em local} damping, 
$M_{CZ}\, u_{rms}^{3}/\ld = 1.54\ M_{CZ}\, u_{t}^{3}/\ell_{D}$. 
The condition of global balance between all driving and damping terms
gives a value for $\ell_{D}$, but the local value of $u_{t}$ is used. 
This gives a relatively good, smoothed fit to the actual inferred damping 
throughout the turbulent convection zone, and departs only in the boundary
layers where the velocity field is due primarily to gravity waves.
{\em The net effect of the remaining terms in the KE 
equation is to modify the velocity field so that the damping
is more smoothly distributed over the turbulent region than 
is the driving.}

The agreement in Figure~\ref{fig99a}, right panel, between the actual 
inferred dissipation and our estimate provides strong support for our
introduction of the ``Kolmogorov dissipation'' in both its global and
local forms.

Fig.~\ref{fig99b} shows the spatial behavior of the flux divergence 
terms, $\bf \nabla \cdot F_{K}$ for flux of kinetic energy and
$\bf \nabla \cdot F_{P}$ for the pressure correlation
flux. These terms have
significant positive and negative contributions, 
which cancel upon averaging over radius, and therefore are
more important than Table~\ref{tablex} would suggest. The change of
sign in a divergence implies that these terms move kinetic energy.
In particular, they remove kinetic energy from the region where
buoyancy is strong, and add it to regions where the buoyancy
is weak. 
The divergence of the kinetic energy flux (Fig.~\ref{fig99b}, left
panel) is the most significant in transporting energy. It moves kinetic
energy from the region in which buoyancy driving exceeds the inferred
damping, and toward the convective boundaries.

In Fig.~\ref{fig99a}, left panel, we saw that there
is negative buoyancy at the convective boundaries; this is due to 
buoyancy braking of the convective motion. The pressure correlation flux
(Fig.~\ref{fig99b}, right panel)
is most effective in the deficit regions right at the convective 
boundary, and generates elastic response (waves) in the stably stratified
regions outside the convection zone.
This two-step behavior, with first $F_{KE}$ and then $F_{P}$ carrying
energy to the edge of the convection zone, is mirrored at
the upper convective boundary, but at lower amplitude.

The $p' \nabla \cdot u'$ term, shown in the left panel of 
Fig.~\ref{fig99c}, does the same sort of thing at a much reduced level.
In the right panel is shown the time derivative of the kinetic 
energy, which is small. The convection is close to a steady state
behavior.

There is a simple picture which explains these trends. (1) The extent
of turbulence is 
limited by energy and by boundaries. In a star, turbulence will mix
even stably stratified layers so long as sufficient kinetic energy
is available to supply the work necessary. (2) Turbulence takes 
ordered motion on the large scales and converts it to disordered 
motion on small scales. It makes the flow more isotropic. Kolmogorov
dissipation is derived with the assumption of 
homogeneity and isotropy, and so becomes
a better approximation as turbulence acts. 

The convective motion is driven by buoyant acceleration, parallel to
the gravitational acceleration vector, and is necessarily large-scale 
and anisotropic. The large-scale order of this motion is 
destroyed by turbulence, which spreads through all accessible regions.
Thus, over all the convective region, except right at the boundaries,
the flow is made more isotropic, and the Kolmogorov damping becomes 
better approximated by the local isotropic expression 
\begin{equation}
\label{lkolgeps}
\varepsilon = \rho \ubar^{3}/\ld,
\end{equation}
(per unit volume).
After carefully distinguishing between global Kolmogorov damping 
(Eq.~\ref{kolgeps}) and the local version (Eq.~\ref{lkolgeps}), 
we find that turbulence tends to drive fluxes in such a way as to 
make both valid.

\subsection{The Phase Shift between Damping and Driving}

\begin{figure}
\figurenum{5}
\plotone{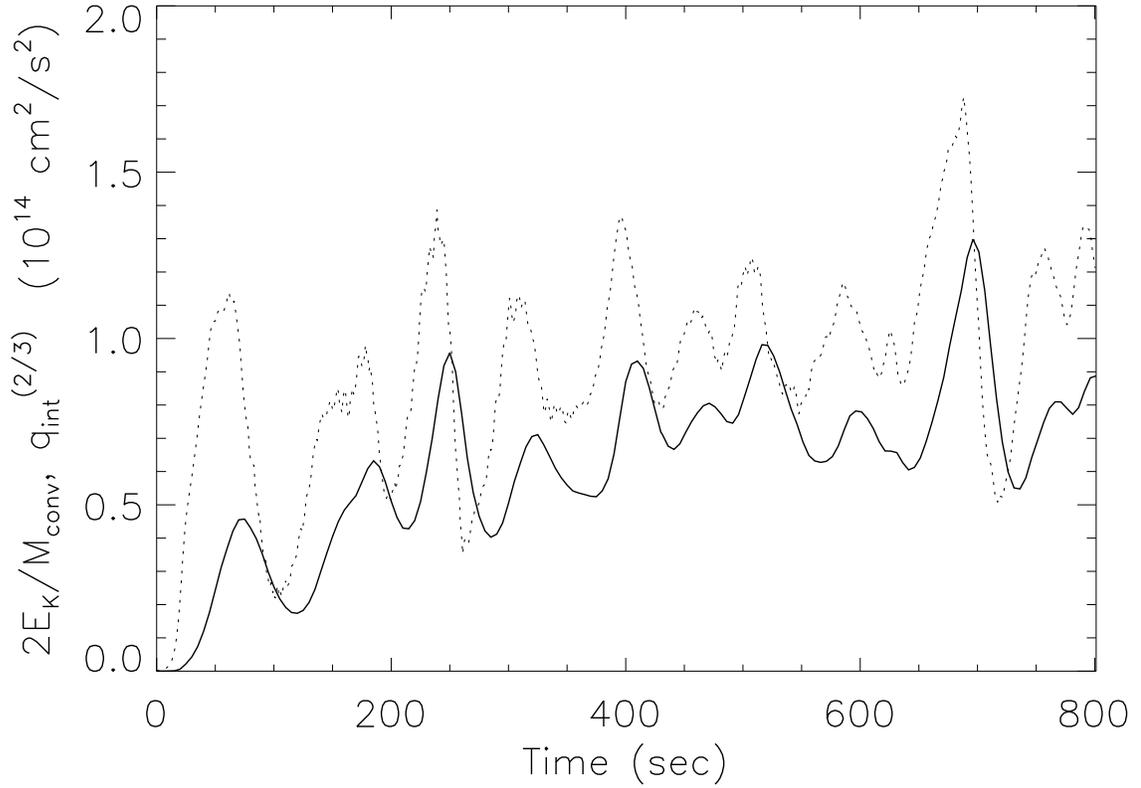}
\caption{Turbulent velocity ${\vrms}^{2}=2 E_{turb}/M_{CZ}$ (solid) and 
$q_{int}^{2/3}$ (dotted) in the 
convection zone, versus time. Integral buoyancy flux is 
$ q_{int} =\int_{CZ} q dr $, and has units of velocity cubed.
Power spectra for both variables peak at 89 seconds; a transit time is $51
\rm\ s$. The kinetic energy lags the buoyant flux by roughly 
20 seconds. The average turbulent kinetic energy in the convection
zone is about $0.64 \times 10^{47}\rm \ ergs$, with significant 
fluctuations about that value.}
\label{fig5}
\end{figure}

Having explored the spatial dependence of the KE equation, we now 
examine its time dependence. The flow, while wildly fluctuating, has
an orderly statistical behavior. To see this, we examine the behavior
of kinetic energy, integrated over the volume of the convection zone,
as a function of time. The integral buoyancy flux, $q_{int}=\int 
\langle {\bf g \cdot u' }{\rho' \over \rho} \rangle dr$, has 
dimensions of velocity cubed.
It is convenient to plot kinetic energy and $q_{int}^{3/2}$ on the 
same graph for detailed comparison. Figure~\ref{fig5} shows the 
time behavior of damping and driving terms.
There are two types of time dependence: (1) a set of bursts 
(pulses), and (2) a secular increase in amplitude (related to the
lack of total balance between nuclear heating and neutrino cooling
over the convective zone (\cite{arnett}, see Ch.~10). 

For a static steady state, these two curves would be almost identical.
If for simplicity we assume a planar convective region and 
neglect the smaller terms in the KE equation (Eq.~\ref{a12}) in favor of
buoyancy and damping, we find the simple result
$ {\vrms}^{3}/\ld \approx { 1 \over \lcz}\int q dr$, 
or $ \ld /\lcz \approx {\vrms}^{3} / \int q dr  $.

The power spectra of both curves have a peak at 89 seconds.
Both exhibit strong fluctuations;
the buoyancy flux precedes the kinetic energy by roughly 20 seconds. 
{\em We identify\footnote{In implicit large eddy simulations (ILES)
like this, the numerical cascade only goes down to the grid scale.}
this lag with the
time it takes for the turbulent cascade to react to changes in the 
flow.} This is suggestively close to our previous estimate of the
turbulent decay time of about 14 seconds.

Buoyancy is primarily a property of the largest scales,
while damping is a property of the smallest. The large separation of 
scales means that they are not tightly coupled (except on average), a 
characteristic of turbulent systems.

The buoyancy flux reaches a high value before turbulence can 
stop its rise, so that it overshoots the steady state condition. 
This leads to excessive velocities, which then cascade to excessive
damping. In our simulations, the lag in dissipation behind buoyant
driving
aids fluctuations about the nominal steady state condition.

This ``boom and bust'' cycle 
is reminiscent of prey-predator relations and the 
logistic map \citep{may}, which also have chaotic behavior.
We find that an iterated map of the delay
in turbulent damping is quantitatively inadequate to drive the
fluctuations. They are driven by nuclear
burning, as our simulation (described below) of turbulent decay implies.
It will be interesting to explore
whether this fluctuating behavior depends upon resolution 
(we expect it to be 
dominated by the largest scale eddies, and weakly dependent on
resolution).

Consider the average conditions from 200 to 800 seconds in the 
simulations. The average level of turbulent
kinetic energy is about 0.7 of $q_{int}^{2/3} $.
If we ignore the smaller terms in the KE equation,  we find
\begin{equation}
    \alpha_{D}= \ld/\lcz = 1.54\ u_{t}^{3} \Big / \int q dr  
    \approx 0.89.
\end{equation}
This is roughly what we get from Table~\ref{tablex}, so
we conclude that to our accuracy, {\em the dissipation length 
in our simulations is roughly 
the depth of the convective zone}, consistent with Kolmogorov 
theory. The appearance of $\ell_{CZ}$ here is due to integration over the
convection zone, not to any assumption about the nature of the 
largest eddies.

\subsection{The Decay of Turbulence}

\begin{figure}
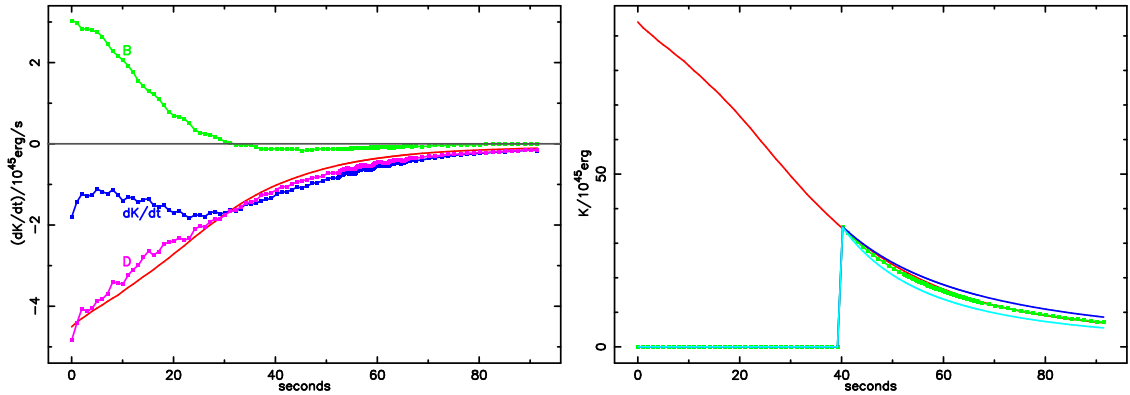

\figurenum{6}
\includegraphics[angle=270,scale=0.3]{f6a.eps}
\includegraphics[angle=270,scale=0.3]{f6b.eps}
\caption{(left panel) The rate of decay of turbulent kinetic energy after oxygen
burning
and neutrino cooling are turned off. The dominant terms are shown: 
$B= \int W_B \dV$ is the
increase of kinetic energy due to buoyant driving,
$dK/dt$ is the
time derivative of the total kinetic energy in the convection zone,
and
$D=\int \varepsilon_K \dV$ 
is the inferred dissipation from the turbulent spectrum
($dK/dt-B$).
We assume that little kinetic energy escapes the convection zone. 
The solid line shows the damping rate approximated by
$\varepsilon_K = - M_{CZ} \, {\vrms}^3/\ld $ 
with $\ld = 4.35 \times 10^8 \rm\ cm$.
After 30 seconds, the buoyancy has died away but damping remains.
(right panel) Kinetic energy as a function of time (solid line), 
and fits of the damping after 40 seconds for 
$\ld/ 10^8 {\rm\ cm} = 2.1$, $2.6$ (best) and $3.1$. 
In this phase there is no driving to create anisotropy, and the
flow becomes more isotropic on average.
}
\label{fig6}
\end{figure}

The decay of turbulence is a complex topic  \citep{batch}. 
Here we will utilize it to provide independent
estimates of the damping length. These estimates will differ in detail
from those of driven convection because the flow details change, and
the dissipation is a function of the flow properties. Nevertheless
the sizes of the damping lengths are found to be comparable.

In the left panel in Fig.~\ref{fig6}, we show an independent 
simulation: the rate of decay of turbulent kinetic 
energy, after oxygen burning and neutrino cooling are artificially 
turned off at 439 seconds (in the middle of the previous simulation).

If we neglect the small terms for energy flux escaping the convective
zone, we may express the kinetic energy equation (Eq.~\ref{a12})
more concisely as
$D = dK/dt - B $, where $D$ is the inferred dissipation.   
After 30 seconds, the buoyancy driving term $B$
becomes small, and $dK/dt$ tracks the turbulent dissipation $D$. 
Notice that $dK/dt$ is slightly below $D$, possibly because 
of entrainment of stable matter at the convective boundary which 
requires energy. If we globally fit the inferred damping term by 
$\varepsilon_K = - M_{CZ} \, {\vrms}^3/\ld $, we have 
$\ld = 4.16 \times 10^8 \rm\ cm = 0.97\ \lcz$. 

The right panel in Fig.~\ref{fig6} shows a fit to the kinetic energy,
starting 40 seconds after the burning was switched off, and the 
fossil buoyancy had died away.
The kinetic energy may be represented analytically if $dK/dt = D$,
so that this late part of the damping curve is reasonably well approximated
by a lower damping length, $\ld = 2.6 \times 10^8 \rm \ cm = 0.61\ \lcz$.

This independent simulation supports the 
identification of sub-grid damping with isotropic Kolmogorov damping,
and the representation of damping by the global expression
(Eq.~\ref{kolgeps}), 
and $\ld  \approx (0.6 {\ \rm to \ } 1.0) \lcz$.

\begin{deluxetable}{rr}
\tablecaption{Decay Lengths for Different Flows\label{tableld}}
\tabletypesize{\small}
\tablewidth{260pt}
\tablehead{
	\colhead{Flow} & \colhead{$\alpha_{D}=\ell_{D}/\ell_{CZ}$}
}
\startdata

Quasi-steady Oxygen Burning  & 0.89 \\
Fossil buoyancy in decay     & 0.97 \\
Pure decay                   & 0.61 \\
\enddata
\end{deluxetable}

These and the previous estimates for $\ld$ seem to vary beyond 
the statistical 
accuracy of the analysis. Replacing the complexity of turbulent
dissipation by the simple expression 
$\varepsilon_K = - M_{CZ} \,  {\vrms}^3/\ld $, with constant
$\ld$ is a vast change. 
This quantity $\ld$ is a property of the flow,
and indirectly of the problem being addressed.
In particular, the pure decay is more isotropic than the driven cases
\citep{ogilvie}.
As Table~\ref{tableld} indicates, the ratio $\ell_D / \ell_{CZ}$ is 
different for different flows, but of order unity.
Further invesitgation with different
simulations and different physical parameters is needed to understand
more precisely the general behavior of $\ld$.

\section{General Equations for Stellar Evolution}

How is this approach related to the standard equations of stellar
evolution?  

\subsection{Internal Energy Equation}
In MLT a connection is made between superadiabatic 
gradient $\Delta \nabla$ and convective velocity. This allows the
convective enthalpy flux to be expressed in terms of $\Delta \nabla$
(see \cite{kippen}, \S~7, \cite{clay83}, \S~3-5, and \cite{hansen}, 
\S~5). The turbulent kinetic energy equation (Eq.~\ref{a12}) can
perform
the same function, if the convective velocity is identified with the
rms velocity $\urms$ at a given radius, and used in the enthalpy 
flux in the same way. This replaces the parameterization used in 
MLT with a physical constraint.

We need to rewrite the total energy equation (A.6 in 
\cite{ma07b}) in the form of the internal energy equation used in
stellar evolution. To get the internal energy equation, we subtract
the kinetic energy equation (A.12) from the total energy equation 
(A.6), and find
\begin{eqnarray}
\partial_{t} \langle\overline{ \rho E_{I}}\rangle +  
{\bf \nabla \cdot }\langle \overline{ {\bf u_{0}} 
(\rho E_{I} 
+ p_{0} )}\rangle =\nonumber \\
- \nabla \cdot \Big [{\bf F_{I} + F_{r} }\Big ] 
- \langle\overline{ p' \nabla \cdot {\bf u'} }\rangle \nonumber\\
+ \varepsilon_{K} + \langle\overline{ \rho_{0} {\bf u_{0} \cdot g 
}}\rangle
+\langle \overline{ \rho \epsilon}\rangle.
\end{eqnarray}
The left-hand side is just $\rho(dE_{I}/dt + p_{0}dV/dt)$ in
lagrangian
(co-moving) coordinates (\cite{ll59}, see Ch.~1), defining $\rho_{0} = 
1/V$, so that
\begin{eqnarray}
d\langle \overline{E_{I}}\rangle /dt + \langle \overline{p_{0} dV/dt
}\rangle 
= & \nonumber \\   
- { 1 \over \rho_{0}}{\bf \nabla \cdot } \Big [ {\bf F_{e} + F_{r}
}\Big ] 
+\langle \overline{\epsilon} \rangle \nonumber\\
- \langle\overline{{ p'\over \rho_{0}}{\bf \nabla \cdot u' }}\rangle
+ \varepsilon_{K}/\rho_{0}  \nonumber\\
+ \langle\overline{ {\bf u_{0} \cdot } ( {\bf g }
-{1 \over \rho_{0} }{\bf \nabla }p_{0}) }\rangle
\label{eq-ei}
\end{eqnarray}
The first two lines are the usual terms. 
Here ${\bf F_{r}}$ is
the heat flux due to radiative diffusion, and in this frame
the flux of internal energy 
${\bf F_{I}}$ becomes the enthalpy flux carried by convection
${\bf F_{e}}$ (see \cite{tennekes}, p.~33, and our Eq.~\ref{fenthalpy}).
The nuclear heating term $\epsilon$ includes neutrino cooling.

We will rewrite Eq.~\ref{eq-ei} in more familiar notation (see 
\cite{arnett}, \cite{clay83}, \cite{hansen}, or \cite{kippen}).
\begin{eqnarray}
dE/dt + PdV/dt = -{1 \over \rho}{\bf \nabla \cdot [ F_{e} + F_{r} ] }
+ \epsilon \nonumber \\
- \langle\overline{ { p'\over \rho_{0}}{\bf \nabla \cdot u' } 
}\rangle + \varepsilon_{K}/\rho_{0}  \nonumber\\
+ \langle\overline{ {\bf u_{0} \cdot } ( {\bf g }
-{1 \over \rho_{0} }{\bf \nabla }p_{0}) }\rangle
\label{eq-ei2}
\end{eqnarray}
The first line contains the usual formulation. The new terms are:
$ \langle\overline{{ p'\over \rho_{0}} {\bf \nabla \cdot u'} }\rangle$
which represents the compressional work done by pressure fluctuations
(which also appears in the kinetic energy equation and we have seen to
be small here), 
$\varepsilon_{K}/\rho_{0}  $ which is the deposition of heat by the
Kolmogorov turbulent cascade, and 
$\langle\overline{ {\bf u_{0} \cdot }( {\bf g} -{1 \over \rho_{0} 
}{\bf \nabla} p_{0}) }\rangle$, which is zero in hydrostatic equilibrium
without rotation or expansion, and
drives meridional circulation for rotating, radiative stars
\citep{tas,clay83}. 

The $\varepsilon_K/\rho_0$ term allows 
turbulent kinetic energy to do dissipative heating, a new
effect not in conventional stellar evolution, and it is not guaranteed to
be negligible. Through Eq.~\ref{a12} it couples the divergence of the kinetic
energy flux (including rotational shear) 
and the wave energy flux to the internal energy.
As an internal energy source term, it can generate entropy and cause
mixing even in radiative regions. For the Sun, this term would give rise
to heating in the photosphere, chromosphere and corona by shocks and wave 
motion (pressure and Alfven waves), for example.

Equations~\ref{a12} and \ref{eq-ei2} represent an extension of turbulent
convection theory for stellar evolution, in which (1) the algebraic relation
for convective velocity and superadiabatic gradient is replaced by a
differential equation (see \cite{spie67}), and (2) the Kolmogorov cascade 
is explicit in the formulation.
Notice that both space and time derivatives appear, making the system
nonlocal and time dependent, unlike MLT. These derivatives allow the treatment
of boundary dynamics in a physical way. We will explore specific 
implementations in a subsequent paper, including entrainment at 
convective boundaries.

Let us now consider some simplifications in order 
to clarify the meaning of these equations.
In particular, we assume time invariance (so the time derivatives are small on
average), no overall background motion (${\bf u_0} = 0 $), 
and little work done by pressure perturbations on the velocity perturbations.
As shown previously, these are reasonable approximations except in 
extreme cases. Then Eq.~\ref{a12} becomes
\begin{equation}
\overline{ \langle \rho' {\bf g \cdot u'}\rangle} - {\bf \nabla \cdot 
(F_K +  F_P ) } = \varepsilon_K,
\end{equation}
and Eq.~\ref{eq-ei2} becomes
\begin{equation}
\rho \epsilon  - {\bf \nabla \cdot ( F_e + F_r) } = - \varepsilon_K 
\end{equation}
The coupling of internal energy and turbulent kinetic energy occurs through
the Kolmogorov cascade, which creates internal energy  by damping kinetic energy.
Eliminating $\varepsilon_K$, we have
\begin{equation}
\rho \epsilon = - \overline{ \langle \rho' {\bf g \cdot u'}\rangle}
+  {\bf \nabla \cdot ( F_e + F_P + F_r + F_K) }. 
\label{eq-rhoe}
\end{equation}
In a steady state, the nuclear energy generation must balance the divergence
of fluxes out of the region and supply the work needed to maintain the convective
flow. This is different from the usual formulation used in stellar evolution
in that there are new terms 
($- \overline{ \langle \rho' {\bf g \cdot u'}\rangle} + 
{\bf \nabla \cdot ( F_K + F_P) } $), 
which combine to equal $\varepsilon_K$, the dissipation
rate due to the Kolmogorov turbulent cascade. If the turbulent velocity is
nonzero, these corrections are also nonzero, leading to the conclusion that
the standard formulation of stellar evolution is wrong in neglecting turbulent
heating. The motion implied by convection will carry kinetic energy, 
drive pressure and gravity waves into radiative regions, and give local
microscopic heating as it dissipates in a transit time, effects that 
should no longer be ignored in stellar evolution. Through its 
dependence on $\zeta$, for a given convective enthalpy flux, the kinetic 
energy flux is dependent upon the equation of state. 

For a low Mach-number flow and radial coordinates,
\begin{equation}
 \overline{ \langle \rho' {\bf g \cdot u'}\rangle}
= \beta_T g F_e / C_P T,
\end{equation}
so that $ {\bf \nabla \cdot F} \rightarrow dL/dm$,
then Eq.~\ref{eq-rhoe} becomes
\begin{equation}
\epsilon = -{\nabla_a \over m_p}L_e + {d \over dm}( L_e + L_P+ L_r + L_K),
\label{eq-rhoe2}
\end{equation}
where $ \nabla_a = \beta_{T} P V/ C_P T$, $m_p = 4 \pi r^2 \rho H_P$, and
$H_P$ is the pressure scale height. For an ideal gas, $\nabla_a
\rightarrow (\gamma -1)/\gamma $, or 0.4 for $\gamma = 5/3$.
For gases with a specific heat at constant pressure which is large
compared to the specific heat at constant volume (such as partially
ionized plasma or electron-positron plasma), $\nabla_a$ is
smaller (see below).

\section{Comparison to Other Simulations}

\begin{deluxetable}{rrrrrrrr}
\tablecaption{A Comparison of Parameters from
some 3D simulations\label{table1b}}
\tabletypesize{\small}
\tablewidth{400pt}
\tablehead{
	\colhead{Reference} & \colhead{$\lcz/H_{P}$}
 & \colhead{ $\alpha_{T}$ }
 & \colhead{$\alpha_E$} 
 & \colhead{$\alpha_v$} 
 & \colhead{EOS}
 & \colhead{Bnd.}
 & \colhead{Zones}
}
\startdata
\cite{ma07b}  & 2.0 & 0.73 & 0.70 & 1.22 & $e$-pair & yes  & 4.0(6)\\
\cite{pw00}   & 4.5 & 2.04 & 0.80 & 2.70 & ideal    & grid & 6.7(7)\\
\cite{cs89}   & 4.8 & 1.05 & 0.83 & 2.16 & ideal    & grid & 3.6(4) \\
\cite{kim95}  & 6.0 & 1.42 & 0.85 & 2.16 & ionize   & yes  & 3.3(4) \\
\cite{cs96}   & 6.8 & 1.30 & 0.68 & 2.60 & ideal    & yes  & 4.8(5) \\
MLT           &   & $\alpha/2$  & 1.0  & $\alpha/\sqrt{2}$ &    & &  \\
\enddata
\end{deluxetable}

These theoretical ideas may be tested by 
application to other simulations of turbulence. 
Several groups have compared 3D simulations to MLT predictions
\citep{cs89,cs96,kim95,kim96,ma07b,pw00,pwj00}, and published
sufficient detail to allow easy quantitative comparison.
All  agree that MLT is somewhat successful, but derive different
values for some of the MLT parameters. This suggests that
these parameters may not be universal, but a function of the conditions
of the simulated flow, and that our theoretical analysis may be
able to put them on a common basis. See also \cite{abbett97,lfs99,
tramp99,brand05}, who have also compared simulations to MLT.

These simulations are not a homogeneous set, so that global
comparisons on this data set must be taken with caution. 
\cite{pw00} and \cite{ma07b}
used PPM codes while the Yale group \citep{cs89,cs96,kim95,rob04} used 
compressible viscous codes with sub-grid scale modelling (and much lower
resolution). \cite{cs96} and \cite{ma07b} had the convection zone 
bounded by stable regions while the others did not; these boundary 
conditions give rise to new phenomena \citep{ma06,ma07a}. The deeper 
convection zones developed more
anisotropic flows, and velocities approaching the sound speed
(e.g., \cite{catt91,wpj03}).
\cite{pw00} carefully attempted to compensate for these effects,
the Yale group used damping to tame them, and the soft equation of
state and shallower depth made them small for \cite{ma07b} (see below).
\cite{pw00} found that they needed to shift the apparent mixing length
by about 30 percent, down from $\alpha = 3.53$ to $\alpha = 2.68$. 
The simulations of \cite{ma07a} included an oxygen-burning shell 
(with an electron-positron equation of state, see Table~\ref{table2}), 
and those of \cite{kim95} and \cite{rob04} included a solar photosphere (with 
strong ionization effects in the equation of state). Our simulations 
are driven by nuclear heating at the bottom of the convection zone; 
the others are drivn by cooling at the upper boundary (like the Sun).

Table~\ref{table1b} summarizes the inferred parameters. The
entries are ordered in increasing depth of the convective zone as
measured in pressure scale height units ($\ell_{CZ}/H_P$), which corresponds
to increasing asymmetry in the vertical direction. 
The choice of equation of state (ideal gas, $e^-e^+$-pair gas, ionized
plasma) and of the boundary conditions
affect the simulations. Even the definition of the
depth of the convective zone may be modified depending on whether
the grid includes the stable bounding region ("yes") or not ("grid"). 
The bottom line in the table gives the traditional MLT values for
several parameters \citep{kippen}. The last column gives the number of zones on
the computational grid.

\subsection{Convection Parameters}

We now examine each of a set of important convection parameters 
(see \cite{pw00} and \cite{ma07b} for details). These parameters
reflect the various uses of the MLT parameter $\alpha$, and are
a convenient and concise way to compare the simulations of 
compressible turbulent convection.

\subsubsection{$\alpha_T$}

\begin{figure}
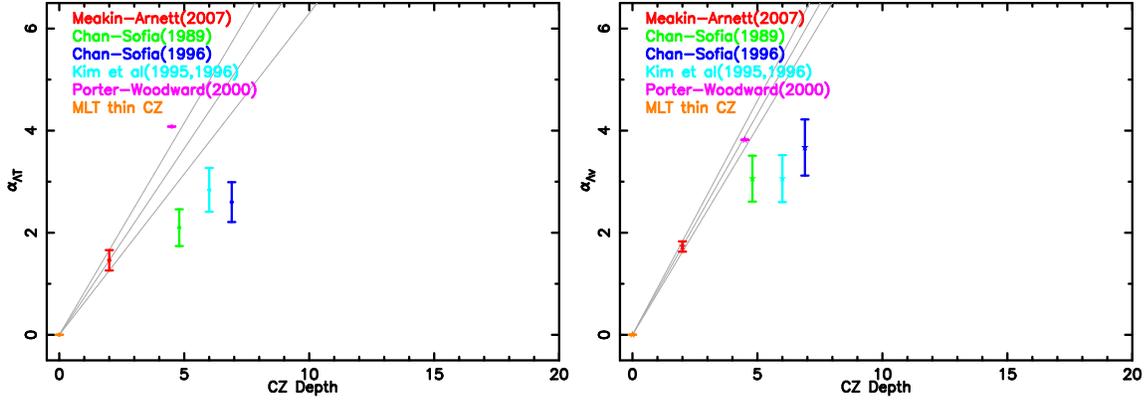

\figurenum{7}
\includegraphics[angle=270,scale=0.3]{f7a.eps}
\includegraphics[angle=270,scale=0.3]{f7b.eps}
\caption{Predicted $2\alpha_{T}=\alpha_{\Lambda,T}$ (top panel) and
$\sqrt{2}\alpha_{v}=\alpha_{\Lambda,v}$ (bottom panel),
 as a function of depth of the convective zone in units of
$H_P$. This scaling \citep{ma07b} gives ``alphas'' which are
comparable to the MLT values and each other.
Some of the error bars are large; new simulations are needed to
determine how well such results follow a single curve 
(Meakin \& Arnett, in preparation). 
In the limit of small depth, the mixing length must be no larger than 
the depth itself; hence the point at zero depth is an analytic result. 
The \cite{ma07b} and \cite{pw00} points and the origin are close
to colinear.
It appears that both $\alpha_{T}$ and $\alpha_{v}$ are  functions of 
depth of the convective region, and not universal constants.
}\label{fig7}
\end{figure}

In 3D simulations, 
a correlation is found between rms temperature fluctuation 
$\Tbar = \Trms$ and the superadiabatic gradient 
$\Delta \nabla = \nabla - \nabla_a -\nabla_x$,
using the conventional notation in astrophysics of $\nabla = \partial 
\ln T / \partial \ln P $ \citep{kippen,hansen,clay83}.
Here, $ \nabla_{a} = \partial \ln T / \partial \ln P$, taken at 
constant entropy and composition, and $\nabla_{x}=-\nabla_{\mu}$ 
is the remaining part due to possible compositional change.
Then,
\begin{equation}
    \Tbar/T_{0} = \alpha_T \Delta \nabla,
    \label{tbart}
\end{equation}
on average, over time and over the volume of the convection zone. 
Consider low Mach-number flow so that 
$\rho'/\rho_0 = - \beta_{T} \  (T'/T_0)$.

In a single convective roll (e.g., \cite{lorenz,tritton}), 
$T'$ is the temperature perturbation amplitude at the horizontal mid-plane. 
In the vertical mid-plane of the roll,
$T' = [(dT/dr)-(dT/dr)_{ad}] \ell_{CZ}/2$ is the corresponding amplitude.
Assuming the amplitudes are comparable,
\begin{equation}
T'/T_{0} = (\ell_{CZ}/2 H_{P})   \Delta \nabla, 
\end{equation}
implying that $\alpha_T \approx \ell_{CZ}/2 H_{P}$, and not a constant. 
In MLT,  $\alpha_T = \alpha/2$, which is a particular
choice for the assumed flow. 

In the simple picture of a single convective 
roll, $ \Tbar/T_{0}$  can give a buoyancy torque, while  
$ \Delta \nabla $
cannot,  because the gravitational acceleration vector is 
directed radially downward \citep{tritton}.
In the \cite{lorenz} model of thermal convection, the difference 
between the two is intimately connected to the onset of chaotic 
behavior.
\cite{ma07b} find $\alpha_{T} \approx 0.73$ (see their Fig.~17), so 
that
\begin{equation}
\alpha_T = 0.73\ \ell_{CZ}/2 H_{P}.
\label{alphaT}
\end{equation}
There is considerable variation in the values of 
$\alpha_T$ shown in Table~\ref{table1b}, with a tendency to increase 
for deeper (more stratified) convection zones.

Notice that for two convective rolls, one atop the other, we would
expect the characteristic roll size to change ($\ell \approx 
\ell_{CZ} \rightarrow \ell_{CZ}/2$, so $\alpha_{T} \rightarrow 
\alpha_{T}/2$, approximately). This explicitly
shows how the convection parameters can be a function of the properties
of the flow itself. 

Figure~\ref{fig7}, left panel, 
shows the behavior of $\alpha_{T}$ as a 
function of the depth of the convection zone, for the computations
listed in Table~\ref{table1b}. The two PPM calculations (\cite{ma07b}
and \cite{pw00}) agree with the scaling indicated in Eq.~\ref{alphaT}.
The calculations using the compressible viscous equations with sub-grid 
modelling for dissipation all lie below Eq.~\ref{alphaT}.
In order to give some idea of the depth needed in simulations of
stellar convection zones, the $x$-axis in Fig.~\ref{fig7}
is marked from zero to
the depth of the solar convection zone (20 pressure scale heights).
None of these simulations describe such an extremely anisotropic case.

\subsubsection{$\alpha_E$ and $\alpha_{K}$}

The enthalpy flux is 
\begin{eqnarray}
    F_{c} = \rho_{0}C_{P} \overline{T'u'} \nonumber \\
    = \alpha_{E} \rho_{0}C_{P} \Tbar\ \ubar ,
\label{fenth}
\end{eqnarray}
where \cite{ma07b} find
$ \alpha_{E} = 0.70\pm 0.03$; in Table~\ref{table1b}, $\alpha_E$ is
relatively constant among the simulations (the total range is about 
20 percent). It is not ruled out that $\alpha_{E}$ 
might be a universal constant, or at least slowly varying. 
Note that $\alpha_E$ is just the correlation coefficient
for $T'$ and $u'$. It seems that $\alpha_{E}$ is not sensitive to the 
Prandtl number $Pr$.  \cite{pw00} have a different $Pr$ (ours is $Pr 
\approx 1$) and get an $\alpha_{E}$ similar to ours.

Using Eq.~\ref{tbart}, this becomes
\begin{equation}
     F_{c} = \alpha_{E} \alpha_{T} \rho_{0}C_{P}T_{0} \ubar
\Delta \nabla. \label{fcudd}
\end{equation}
Similarly, if the kinetic energy flux is
\begin{equation}
    F_{K} = {1 \over 2} \langle \rho u'^{2} u_{z}' \rangle,
\end{equation}
we may define
\begin{equation}
    \alpha_{K} = \langle \rho u'^{2} u_{z}' \rangle / \langle \rho 
    \rangle \langle (u_{z}')^{2} \rangle^{3/2},
\end{equation}
so that $ F_{K} = { \alpha_{K} \over 2} \langle \rho \rangle 
\langle (u_{z}')^{2} \rangle^{3/2}$. Note that the sign of 
$\alpha_{K}$ can be negative.

\subsubsection{$\alpha_v$}

In \cite{ma07b}, 
the correlation between convective velocity (squared) and 
super-adiabatic excess $\Delta \nabla$ is written as
\begin{equation}
\ubar^2 =  (\alpha_v^2 / 4 ) g \beta_{T} H_P \Delta \nabla.
\label{mltvc}
\end{equation}
In MLT, a similar expression is defined, with $\alpha^2/8$ replacing
$\alpha_v^2 /4 $. 
If we have local balance between buoyancy driving and turbulent 
damping, 
\begin{equation}
    \langle \rho' g u' \rangle \approx \langle \rho \rangle 
    \ubar^{3}/\ell_{D}.
\end{equation}
using Eq.~\ref{deltaeq}, Eq.~\ref{tbart}, and  Eq.~\ref{fenth}, we 
have
\begin{equation}
    \ubar^{2} = (\ell_{D} \alpha_{T}\alpha_{E}) \beta_{T} g \Delta \nabla.
\end{equation}
Comparing this to Eq.~\ref{mltvc} we have
\begin{equation}
\alpha_{v}^{2}/4 = \alpha_{T}\alpha_{E} \ell_{D} / H_{P}.
\label{eqalphav}
\end{equation}
Using Eq.~\ref{alphaT} and $\alpha_{E}=0.70$,
\begin{equation}
\alpha_{v} = 1.22\ (\ell_{CZ}/2H_{P})(\ell_{D}/0.9\ell_{CZ})^{{1\over 2}}   
\label{alphav}
\end{equation}
In Table~\ref{table1b} we see that
$\alpha_v$ is variable, and tends to increase with increasing depth
of the convective zone. This is shown explicitly in the right panel
of Fig.~\ref{fig7}. The two PPM calculations are in excellent 
agreement with Eq.~\ref{alphav}, taking $\ell_{D}=0.9\ \ell_{CZ}$
($\alpha_{D}=0.9$), 
while the three compressible viscous calculations lie below it.

Using Eq.~\ref{fcudd}, we have
\begin{equation}
    F_{c} = (\alpha_E \alpha_T \alpha_v /2)
\rho_{0} C_{P} T_{0} \sqrt{g \beta_{T} H_P }(\Delta \nabla)^{3/2}.
    \label{eqfc2}
\end{equation}
Using Eq,~\ref{alphaT},  Eq,~\ref{alphav},
$\ell_{D} = 0.9 \ell_{CZ}$ and $\alpha_{E}=0.70$,
we have
\begin{equation}
\alpha_{E}\alpha_{T}\alpha_{v}/2 = 0.312 (\ell_{CZ}/H_{P})^{2},
\label{3alphas}
\end{equation}
for the factor in Eq.~\ref{eqfc2}.

\subsection{The electron-positron Plasma\label{paireos}}

\begin{deluxetable}{rrrrrrr}
\tablecaption{Thermodynamic parameters for Entropy 
$S/{\cal R}=4.623$\label{table2}}
\tabletypesize{\small}
\tablewidth{320pt}
\tablehead{
\colhead{$T_{9}$} & \colhead{$\rho_{6}$} & \colhead{$\ln P_{max}/P$}
& \colhead{$C_{P}/{\cal R}$} & \colhead{$\beta_{T}$} &
\colhead{$PV/C_{P}T $}
& \colhead{$\nabla_{ad}$}
}
\startdata
2.51& 1.580 & 0.0   & 15.36 & 3.912 & 0.0609 & 0.2382\\
2.31& 1.225 & 0.348 & 14.53 & 3.757 & 0.0638 & 0.2397\\
2.11& 0.930 & 0.724 & 13.62 & 3.592 & 0.0673 & 0.2417\\
1.91& 0.690 & 1.134 & 12.67 & 3.421 & 0.0714 & 0.2443\\
1.71& 0.498 & 1.583 & 11.73 & 3.257 & 0.0762 & 0.2483\\
1.51& 0.348 & 2.080 & 10.83 & 3.106 & 0.0814 & 0.2529\\
\enddata

\end{deluxetable}

The thermodynamic character of the electron-positron plasma 
in the oxygen burning shell is significantly different from an ideal 
gas, an effect which must be taken into account in comparisons to
simulations which use different equations of state. 
This is illustrated in Table~\ref{table2}; we use the 
Helmholtz equation of state of \cite{ts00}. The entropy in the
oxygen burning convection zone is $S/{\cal R} \approx 4.6$ in 
dimensionless units, where $\cal R$ is the gas constant. 
The zone is about 2 pressure scale heights deep
(see column~3). For an ideal gas, the specific heat at constant 
pressure is $2.5\cal R$; $C_{P}/{\cal R}$ is much larger for the
plasma,
ranging between 10 and 16 (column 4). 
Column~5 gives the value of $\beta_{T}$, which is
unity for an ideal gas, but lies between $3$ and $4$ for the 
plasma.
Column~6 gives the ratio of
$ PV / C_{P} T = p_{0}/\rho_{0}C_{P}T_{0}$, which is 0.4 for an ideal
gas.
 The ratio of the buoyancy flux to the enthalpy flux 
(Eq.~\ref{fetoby})
is proportional to $\nabla_a = \beta_{T} PV/C_{P}T$. The same factor appears
in the ratio of the kinetic energy flux to the enthalpy flux. For the 
ideal gas $\nabla_a = 0.4$, but is smaller for the plasma (column 7).
For the same convective enthalpy flux, the pair-plasma has a lower
kinetic energy flux, so the velocity scale is lower.

\subsection{Direction of the Kinetic Energy Flux}
The kinetic energy flux is averaged over angle at a given radius, and
averaged over two convective turnover times. As Fig.~\ref{fig5} shows, 
this time span covers significant dynamic behavior. Roughly speaking,
an unstable configuration forms, becomes dynamic, reforms, becomes
dynamic again, and so on. Over this turnover timescale the nuclear
burning provides the energy necessary to drive the turbulent kinetic
energy 
($ \epsilon \tau \approx {1 \over 2}\langle\overline{u'^{2}}\rangle $).

The convective instability is closely related to the Rayleigh-Taylor
instability \citep{ch61}, which has been produced dramatically in high
energy-density plasma experiments \citep{rem99}, i.e., under star-like 
conditions and well into the nonlinear growth regime. The 
characteristic behavior is the rise of mushroom-shaped higher entropy
plasma and the fall of spike-shaped lower entropy plasma. If we
consider a closed volume, these motions are accompanied by slower
return motions which maintain mass conservation. The kinetic energy
flux scales with velocity cubed, and so is dominated by the fast
moving mushrooms and spikes. In a symmetrical system we will have
an upward kinetic energy flux (from the mushrooms) in the region 
above the horizontal midplane of the volume, 
and a downward kinetic energy flux in the region below.
If we average over several cycles (turnover times) the kinetic energy
flux with be dominated by these motions, being positive (upward) above
the midplane and negative below.
This is qualitatively similar to the simulation results of \cite{ma07b}
(see their Fig.~21 and Fig.~22).

This simple picture is complicated by an up-down asymmetry due to
stratification (hydrostatic structure). The depth of the convection 
zone in pressure scale heights, 
$\ell_{CZ}/H_{P} = \ln ( P_{bot}/P_{top})$, 
is a direct measure of
the up-down asymmetry. Upward flows move into regions of lower 
pressure, and expand; downward flows move into regions of higher 
pressure, and are compressed. The smaller area of the downflows
implies higher velocities relative to a coordinate frame 
containing the convection zone
(a lagrangian frame). This will favor
downward (negative) kinetic energy fluxes. The neglect 
of kinetic energy fluxes \citep{ebv3} corresponds to the limiting case
of a shallow convection zone. For deep convection zones, there is
a strong bias in favor of fast downward plumes \citep{nordstein,sn98}, 
and these dominate the flow for simulations with 
$\ell_{CZ}/H_{P} \ge 4$ or so. In \cite{ma07b}, which has a depth 
$\ell_{CZ}/H_{P} =2$, the convective zone is skewed, so that the
surface, which separates the positive and the negative kinetic energy
fluxes, moves nearer to the bottom of the convection zone. For 
\cite{pw00}, where $\ell_{CZ}/H_{P}$ is larger,
the positive kinetic energy flux is overwhelmed, and
the direction of the kinetic energy flux is opposite to that of the
much larger enthalpy flux. We expect this to be a general property
of deep convection zones.

There is a natural limit to the depth of convection zones expected
in active burning regions. Entropies in active burning regions vary
relatively slowly (\cite{arnett}, Ch.~10), so that the depth of a
convection zone implies a value of the temperature ratio between 
bottom and top. In Table~\ref{table2}, the convection zone extends 
down almost to neon burning
temperatures ($T \approx 1.5 \times 10^{9}\rm \ K$). Further extension
will entrain new fuel into the oxygen convective shell, which will 
burn at the {\em top} of the convection zone, choking the flow.
For the last stages (C, Ne, O and Si burning), 
the burning temperatures for different fuels are fairly close 
together, implying that their related convection zones will tend to be 
relatively shallow. They are also close together in radius, raising the issue of 
interactions between them \citep{ma06}. 

Here 
$ \ell_{CZ}/H_{P} =\ln P_{bot}/P_{top} = {\gamma \over \gamma 
-1 } \ln T_{bot}/T_{top}$. 
For example, for advanced burning stages
or radiation dominated regions, $\gamma \approx 4/3$, and
$\gamma/(\gamma -1) = 4$. For helium and
hydrogen burning, $T(He)/T(H) \approx 13$ so 
$\ell_{CZ}/H_{P} \le 4 \ln 13 \approx 
10$. A helium burning convective zone will not be deeper than about 10
scale heights unless the overlying matter is devoid of hydrogen.
For carbon burning the corresponding depth is $\ell_{CZ}/H_{P} 
\la 3$, unless devoid of H and He fuel.
Surface convection zones may extend down to the hydrogen burning
regions. Very roughly, $\ln T(H)/T_{e} \approx 8$, so
$\ell_{CZ}/H_{P} \la 32$, using the structure of an $n=3$ polytrope.

\subsection{Magnitude of the Kinetic Energy Flux}

\begin{deluxetable}{rrrr}
\tablecaption{Comparison of 
some 3D simulations\label{table1a}}
\tabletypesize{\small}
\tablewidth{320pt}
\tablehead{
\colhead{Reference} & \colhead{$\lcz/H_{P}$} 
 & \colhead{ ${\nabla_a }$ }
 & \colhead{$F_{K}/F_{c}$} 
}
\startdata
\cite{ma07a}  & 2 & 0.24 & -0.018\\
              &   &      &       +0.014 \\
\cite{pw00}   & 4.5 & 0.40  & -0.3 \\
\cite{catt91} & 5 & 0.40  & -0.35\\
\cite{cs89}   & 4.8 & 0.40  & -0.35\\
\cite{cs96}   & 6.8 & 0.40  & -0.50\\
\enddata
\end{deluxetable}

Table~\ref{table1a} gives the ratio of kinetic energy flux to
enthalpy flux for several 3D simulations. This ratio is much smaller
in our simulations (by a factor of 35 to 50) than in the others,
and $F_{K}$ changes sign in our convection zone.

As in \citep{ma07b} we can write the KE to enthalpy flux ratio according 
to mixing length relationships,

\begin{equation}
\frac{F_{K}}{F_{c}} \sim \frac{\rho v_c^2/2}{\rho c_p T'}\frac{v_c}{v_c}
\sim \left(\frac{\alpha_v^2}{8 \alpha_T}\right) \nabla_{a}
\end{equation}

\noindent and then balance between buoyancy driving and turbulent damping 
through Eq.~\ref{eqalphav} gives

\begin{equation}
\frac{F_K}{F_c}\sim\frac{\alpha_D\alpha_E}{2}\left(\frac{l_{CZ}}{H_P}\right) \nabla_a.
\end{equation}
 
See Eq.~3.14 of \cite{pw00}, which uses a gamma-law equation of state
to generate the sum of kinetic and enthalpy fluxes, and implies an
equivalent flux ratio.

The ratio of $\nabla_{a}$ for the ideal gas to that of
the electron-positron gas gives a factor of 1.6 or so. The
ratios of the depth of the convection zones
give another factor of $\sim$ 5. While these considerations
illustrate the role played by the depth of the convection zone
and go some way towards explaining the trends in KE flux 
it is also important to consider the geometry of the
driving region, which relates to how well a local balance
between buoyancy driving and turbulent damping is achieved as assumed
in Eq.~\ref{eqalphav}.
In particular, the length scale $l_a$ over which buoyancy is imparted 
to the stellar plasma through either heating at the base of the convection zone or
cooling at the top can affect the KE to enthalpy flux ratio dramatically.
This may be understood simply: the flow depends both upon the geometry of the 
convective domain and upon the way in which the fluid is stirred. At present 
we have at least two basic patterns, a mostly negative kinetic energy flux for 
deep convective zones driven by surface cooling (most stellar surface 
convection zones) and a more complex positive-negative convective flux 
for shallower zones driven by nuclear burning. More simulations are underway 
to clarify this issue (Meakin and Arnett, in prep.).

%The kinetic energy flux is 
%\begin{equation}
 %   F_{K} = \alpha_{K} \rho_{0} \ubar^{3} /2,
 %   \label{eqkef}
%\end{equation}
%so the coefficient $\alpha_{K}$ must also change sign.

%Using  Eq.~\ref{eqalphad},  Eq.~\ref{eqalphav},  Eq.~\ref{eqfc2},
%and Eq.~\ref{eqkef},
%\begin{equation}
%    F_{K}/F_{c} = \alpha_{D}\alpha_{E} \zeta \Big ( { 
%    \lcz \over 2 H_{P} } \Big ).
%    \label{fkfe}
%\end{equation}
%See Eq.~3.14 of \cite{pw00}, which uses a gamma-law equation of state
%to generate the sum of kinetic and enthalpy fluxes, and implies an
%equivalent flux ratio.

%The ratio of $\zeta = \beta_{T} PV/C_{P}T $ for the ideal gas to that of
%the electron-positron gas gives a factor of 20 or so. The
%ratios of the depth of the convection zones
%give another factor of 2.5 to 3.5, for a total change of 50 to 70.
%We conclude there is no significant discrepancy in velocity scales
%between the simulations, just a difference in equation of state and
%in convective zone depth.  If we use the
%$\alpha$'s of \cite{pw00} to estimate their flux ratio (their Fig.~11)
%we find better agreement than if we use our values, which supports the
%notion that there is some variation for different flows.

\subsection{Saturation of Kinetic Energy Flux}

Are there limits to the linear rise in energy of convection
that is implied by Fig.~\ref{fig7}? 
Deep convection zones ($\ell_{CZ} \ge 4 H_{P}$) 
have strong negative kinetic energy fluxes.
For very deep zones, the extrapolated kinetic fluxes imply supersonic 
velocities. Such large velocities would generate shock waves,
which would increase dissipation, converting kinetic 
energy into internal energy. The rate scales as the velocity 
difference cubed, which is the same scaling as 
turbulent dissipation in the Kolmogorov cascade \citep{bethe42,boris07}. 
This suggests that even if deeper convection zones did tend to have 
increasingly strong velocities, shock dissipation will become significant, 
and resist the increase of $\ell_{D}$ with increasing $\ell_{CZ}$. 
In this sense, the increase in damping length must ``saturate''  with 
increasing depths. 

Increased damping may come sooner from another 
effect: a change in the nature of the eddies and the size of the damping length.
Physically this would occur as follows: as the deep, fast, narrow 
downflows drive into the convection zone they will give rise to shear 
instabilities at their interfaces, which will lead to mixing with the 
ambient fluid, and exchange momentum through a turbulent viscosity, 
and eventually completely dissolve into the background. This would 
give shorter dissipation lengths, and would begin to
occur before the mach numbers become large enough for significant shock
formation.

We expect the PPM calculations for deeper convective zones to 
``flatten'' in Figure~\ref{fig7} as the Yale simulations do, but at
higher amplitude, due to increased damping with increased depth. 
This hypothesis needs to be tested numerically, which will be 
challenging. Shock waves may cause errors at grid boundaries, deep
convection zones will have longer thermal relaxation times, and maintaining
sufficiently wide aspect angle implies many computational zones for
adequate resolution, for example.

The simulations of the Yale group \citep{cs89,cs96,kim95,kim96,rob04},
which use fewer zones and a strong damping \citep{smag63}, appear to 
have a larger dissipation than the PPM simulations \citep{pw00,ma07b}.
This is qualitatively equivalent to the saturation discussed above,
and may be tested if calculations using the Yale
code (or equivalent) are repeated at higher resolution and/or
lower dissipation.

\section{Some Implications}

\subsection{Waves}
The energy flux terms include both advective transport and waves.
Here we will recall some properties of waves
and their generation (\cite{ll59}, \cite{press81}).
The characteristic frequencies of convective motion are centered
about a frequency $ \omega_{CZ} \approx \vrms / \lcz $.
The convective mach number, $ \ubar /{\sound} = \rho'/\rho_{0}
= p'/\rho_{0} {\sound}^{2}$, is a measure of compressibility
of the flow; here ${\sound}$ is the sound speed.
If the interface between convective and radiative zones moves with
the matter (is volume conserving, on average), 
it generates acoustic waves by dipole emission at a luminosity
$ L_{p-mode} \propto \omega_{CZ}^{6}$ (see \cite{ll59}, \S~73), or
as the Mach number to the sixth power.
For more vigorous motion, the perturbation may give volume changes,
so that the acoustic wave generation by monopole emissivity, or
Mach number to the fourth power. For subsonic flows this
channel is closed to significant energy flow, but open as
$\ubar$ approaches the local sound speed, as it does in the surface
convection zones of many stars, including the Sun, or as it may in
the stage prior to core collapse in massive stars. 

While the exact power dependence may depend upon specific geometries and
degree of interference, a general result seems to be: the gravity wave
channel dominates over the acoustic channel for low mach number flows
(as we observe in our simulations, see \cite{ma07a} for a more detailed
discussion of both the g-mode and p-mode behavior, including mixed 
p- and g- modes).  Both types of waves are generated
by convection interacting with stably stratified bounding regions, 
and the luminosity of each depends upon both the convective
vigor and the impedence at the boundary. Such waves can propagate into stably 
stratified regions \citep{press81,pr81,ya05}.
Generation of gravity waves by convective turbulence has become an 
issue in questions of mixing and angular momentum transport 
\citep{gls91,ct99,tc04,ya05}.

The establishment of a robust estimate of the turbulent velocity
field should improve estimates of wave generation, which has
implications for mass loss, mixing in radiative regions, coronal heating
and helioseismology.
In particular, the pressure correlation flux takes over the energy 
carried by the kinetic energy flux as the convective boundaries are 
approached. This gives a direct connection between the scale of 
turbulent velocity and g-mode wave emission \citep{press81}.

\subsection{Rotation and Magnetic Fields}

A closely similar set of mean-field equations are used in the theory of the
magnetic resonance instability (MRI) in accretion disks 
\citep{balbus,pessah}. If we had included magnetic fields in the
MHD approximation, and assumed strong rotation, our procedure would
have produced an equation for mechanical energy (our Eq.~\ref{a12} 
corresponds to Eq.~17 of \cite{balbus}), and total energy (Eq.~A6 of
\cite{ma07b} to
their Eq.~27). Projection onto a cylindrical coordinate system, 
with the rotational axis oriented parallel to the total angular 
momentum vector, would give an
angular momentum equation (their Eq.~29.). Auxiliary equations provide
for magnetic field dynamics, radiation transport, and nuclear
burning.
This underlying similarity
provides a way to write a more general set of mean-field equations,
of which both stars and accretion disks are limiting cases.
In turn this allows a systematic evaluation of the relative importance
of different effects (rotational mixing and convective mixing, for 
example) which are now considered piecemeal.

Our simulations include the complete set of rotational terms, but
the initial conditions imply that these terms are not exercised except as 
convectively induced shear. Our simulations do not include magnetic
fields in the MHD approximation, but could be generalized to do so 
\citep{stone07,pessah}. Because they interact,  rotation
and magnetic fields should be included together.

\cite{bh94} have argued that in the stellar case, the 
 weak-field MRI dominates over merely hydrodynamic instabilities,
and drives the radiative zones (but not convective zones)
toward solid body rotation.
\cite{hlw00} argued the reverse (based on the H{\o}iland instability
criterion), that convective zones would tend toward rigid body
rotation, and radiative zones would tend to have differential 
rotation. Because the  H{\o}iland instability applies to neutral
fluids, not dense plasma, it is probably not relevant for stars. 
Helioseismology \citep{thom96,howe05,brand07} is showing 
that while the convective
zone of the Sun shows differential rotation, the underlying radiative
zone seems to be tending toward solid body rotation, as the MRI 
arguments suggest. \citep{brown07} have shown that even deep 
convection zones can generate significant magnetic fields, so that
the presence of a stable interface is not necessary for field 
generation.  The rotational state of the central regions of 
the Sun probably depends upon the efficacy of angular momentum
transport by processes related to flow of the plasma, including
g-mode waves \citep{ct99}, as well as magnetic fields.

\subsection{Damk\"ohler Number for Burning}

In general, it is appropriate to decompose the equations in
temperature ($T=T_{0} + T'$) and composition as well. The opacity
and the nuclear reaction rates are often sensitive to both.
\cite{ma06} found flashing due to oxygen burning in vigorous
downdrafts which were fuel-rich (the flashes were too mild to
affect the flow dramatically).
None of these effects are included in
standard stellar evolution theory. For simplicity we supress
this complication for the moment; this means that our opacities and
reaction rates are to be interpreted as averages over fluctuations
in these variables as well. Further
investigation of this issue, with 3D simulations, is desirable.

For this set of simulations, the heating and cooling times are at
least 100 times longer than the transit times, so that we are in
the regime of small Damk\"ohler number $D_m \la 0.01$
\citep{zeldovich,oranboris}.
The release of nuclear energy during a transit time is small relative
to the internal energy in the convection zone, but comparable to the
superadiabatic energy and to the turbulent kinetic energy. The
burning drives the turbulent motions, but only gives moderate pulses
of kinetic energy, as shown in Fig.~\ref{fig5}. This is unlike the
much more complex problem of Type~Ia supernova models, for which 
$D_m$ is large. In our case the necessary averaging over convective
cycles does not seem to be a problem.

This convenient state may not apply to the double shell flash stage
for asymptotic giant branch stars, in which wave driven mixing and
entrainment are likely to complicate the issue of mixing, and 
therefore figure into the question of s-process nucleosynthesis
\citep{bus99,cl08}.

\section{Summary}

We find that our three-dimensional time-dependent simulations of
compressible stellar turbulence (ILES) are well represented by
a master equation (Eq.~\ref{a12}) for kinetic energy which includes dissipation
implied by the Kolmogorov cascade. The damping length is found in three
independent ways, with reasonable consistency, and is the size of
the largest eddy, which is approximately 
the geometric linear dimension (depth) of the convective zone. 
Unlike the mixing length of
MLT, it is not a free parameter, but a mathematical consequence of
the turbulent flow. Balancing turbulent buoyant driving with
Kolmogorov damping provides a reasonable estimate of the damping
length. The divergence of kinetic energy and acoustic fluxes is
nonzero, and provides the mechanism to spread turbulence through
the convective zone, and make the Kolmogorov damping a good 
approximation. The turbulent flow is highly dissipative and must
be maintained by continual driving; this ``frictional heating'' term
is missing from the standard stellar evolution equations, as are
the kinetic energy and acoustic fluxes.

Fluctuations in kinetic energy are significant (of order 50 percent),
and damping lags driving by about a transit time for the convective 
zone.

Comparison with some other simulations, which were dramatically 
different in many respects, gives a consistent picture. 
Turbulent convection is more vigorous for deeper (more stratified) 
convection zones. Turbulent kinetic energies are lower for nonideal
equations of state, such as partially ionized plasmas and 
electron-positron plasmas, to the extent that their specific heat
at constant volume is less that their specific heat at constant 
pressure. The average flow structure changes in a simple way depending
upon the depth of the convection zone; deep convection zones have
downwardly directed flows of kinetic energy, cancelling some of the
upward enthalpy flux.

It appears that extension of this approach, using simulations to define stellar 
convection algorithms, can establish a theoretical model of turbulent
convection that does not require astronomical calibration, but can be 
based upon a combination of computer simulations, terrestrial observations, 
and  experiments. Efforts to implement this general theory in a stellar 
evolution code are underway.

\begin{acknowledgements}
During the course of this project, we lost two friends who were 
leaders in the field of stellar evolution, John Bahcall and Bohdan
Paczynski, to whom this paper is dedicated.
This work was supported in part by  NSF Grant 0708871 and 
NASA Grant NNX08AH19G at the University of Arizona, 
and the ASCII FLASH center at the University of Chicago, 
One of us (DA) wishes
to thank the Aspen Center for Physics and the International
Center for Relativistic Astrophysics (ICRA) for their hospitality, Brian
Chaboyer
for help with the history of the mixing length implementation, Martin
Pessah for discussions of MRI physics, Robert Stein for discussion of 
the effect of the continuity equation on flows, Vittorio Canuto for 
helpful discussions of the philosophy of turbulence modelling,
Martin Asplund for providing
machine-readable copies of solar surface models, and Frank Timmes
for helpful comments on the draft and for providing access to the 
Saguaro computer cluster.
We wish to thank an anonymous referee for insightful comments which
improved both our presentation and our understanding.

\end{acknowledgements}

\appendix

\section{Decomposition}

\par We decompose velocity, density, and pressure fields
into mean and fluctuating components according to

\begin{equation}
  \varphi = \varphi_0 + \varphi',
\end{equation}

\noindent where $\avg{\varphi} = \varphi_0$ and $\avg{\varphi'} = 0$
and
the overbar and brackets indicate time and horizontal averaging, 
respectively. For data handling and analysis purposes we 
consider the fluctuating component of the field to be 
composed of a radial p-mode component and a component 
due to all other sources

\begin{equation}
  \varphi' = \varphi'_p + \varphi'_t.
\end{equation}

\noindent This additional decomposition 
allows us to make a more accurate estimate of the fluctuations
associated with turbulent convection in the presence of
a coherent radial p-mode which is not well sampled in the output
files from the simulation. If the radial p-mode contribution were
well sampled then we would find $\avg{\varphi'_p} = 0$ to 
the degree that the mode is adiabatic.  

\par Consider the radial velocity to be composed of 
the following components

\begin{equation}
  u = u_0 + u'_p + u'_t,
\end{equation}

\noindent where $u_0$ is the mean background expansion, $u'_p$ is the
radial p-mode induced fluctuation, and $u'_t$ is the fluctuation due
to 
turbulent convection and other non-radial modal components.
Averaging,
we find

\begin{equation}
  \avg{u} = \avg{u'_p} + \avg{u'_t} + u_0.
\end{equation}

\noindent Because of our poor sampling of the low order radial p-modes
which have frequencies comparable to the simulation data output rate 
($\delta t \approx 0.5$ s) we find
that the term $\avg{u'_p} > 0$ and contributes a significant error to
the
estimation of $u'_t$. In order to correct for this horizontally
coherent p-mode induced radial displacement, we subtract 
the horizontally averaged radial velocity component at each time step
and estimate the turbulence induced fluctuation by

\begin{equation}
  u'_t = u - \havg{u'_p} - u_0 \approx u - \havg{u'_p}.
\end{equation}

\noindent  The latter approximate equality in the above equation 
is made because of the smallness of the background expansion compared 
to the the r.m.s. velocity fluctuations associated with the turbulent 
convection, $u_0/u'_c \sim 10^{-3}$.

\par The instantaneous fluctuations in pressure and density are
calculated according to

\begin{equation}
  p'_t = p - \havg{p},
\end{equation}
\noindent and,
\begin{equation}
  \rho'_t = \rho - \havg{\rho}.
\end{equation}

\clearpage

\clearpage

\end{document}